\documentclass[prc,letterpaper,twocolumn,showpacs,showkeys,lengthcheck,tightenlines,
nofootinbib,preprintnumbers,superscriptaddress]{revtex4-1}

\usepackage[utf8]{inputenc}

\usepackage{titlesec}
\usepackage{dcolumn}
\usepackage{bm}
\usepackage{amsmath,amssymb,amsfonts}
\allowdisplaybreaks[4]

\usepackage{graphicx}
\usepackage{subfigure}
\usepackage{color}
\usepackage{bbold}

\usepackage{slashed}
\usepackage[svgnames]{xcolor}
\usepackage[english]{babel}
\usepackage{blindtext}
\usepackage{microtype}
\usepackage{tikz}

\usepackage[colorlinks=true,urlcolor=blue,citecolor=blue]{hyperref}
\usepackage{ifpdf}
\usepackage{float}

\graphicspath{{.}{figures/}}

\titlespacing{\section}{5pt}{12pt plus 4pt minus 2pt}{8pt plus 2pt minus 2pt}
\titlespacing{\subsection}{0pt}{12pt plus 4pt minus 2pt}{8pt plus 2pt minus 2pt}

\begin{document}
	
	\title{$\rho$ meson transverse momentum dependent parton distributions }
	
	\author{Jin-Li Zhang}
	\email[]{jlzhang@njit.edu.cn}
	\affiliation{Department of Mathematics and Physics, Nanjing Institute of Technology, Nanjing 211167, China }
	
	\author{Jun Wu}
	\email[]{wujun@njit.edu.cn}
	\affiliation{Department of Mathematics and Physics, Nanjing Institute of Technology, Nanjing 211167, China }
	\affiliation{National Laboratory of Solid State Microstructures, Department of Physics, Nanjing University, Nanjing 210093, China }
	
	
	
	\begin{abstract}
		In this paper, the light-front wave functions (LFWFs) of $\rho$ meson are evaluated in the framework of Nambu--Jona-Lasinio model using the proper time regularization scheme. The transverse momentum dependent parton distributions (TMDs) of the $\rho$ meson are derived from the overlap representations of the LFWFs. We investigate the $\bm{k}_{\perp}$-weighted moments and the $x$-dependent average transverse momentum $\langle k_{\perp}^n(x)\rangle_{\alpha}$ of $\rho$ meson TMDs, $\langle k_{\perp}^n(x)\rangle_{\alpha}$ shows the typical transverse momenta of quark TMDs in our model. Our findings indicate that the average transverse momenta of $\langle k_{\perp}(x)\rangle_{\alpha}$ fall within the range of $[0.3, 0.45]$ GeV, while $\langle k_{\perp}^2(x)\rangle_{\alpha}$ are in the region of $[0.15, 0.30]$ GeV$^2$. Our TMDs and parton distribution function (PDFs) exhibit excellent adherence to positive constraints. The investigation of spin densities within the $\rho$ meson in transverse momentum space reveals axially symmetric distributions for quarks and targets polarized in either longitudinal or transverse directions. Conversely, unpolarized symmetric distributions occur when the quark is longitudinally polarized and the target is transversely polarized, while exhibiting dipolar distortions in the opposite scenario. A comparison between TMDs from LFWFs and those from a covariant approach demonstrates that TMDs from LFWFs more effectively satisfy positive constraints than their counterparts from the covariant approach. Furthermore, the $x$-moments of $\rho$ meson PDFs indicate that $\langle x\rangle_{g_L}$ is larger using LFWFs compared to the covariant approach. Additionally, analysis of quark spin densities within the $\rho$ meson reveals that $\rho_{\uparrow \downarrow}\left(k_x,k_y\right)=f-g_L-\frac{1}{3}f_{LL}$ yields a negative result using the covariant approach, whereas $\rho_{\uparrow \downarrow}\left(k_x,k_y\right)$ derived from LFWFs produces a positive outcome.
	\end{abstract}

	\maketitle
	\section{Introduction }
	The most difficult job of hadron physics is to compute the hadronic transition matrix elements which are fully governed by non-perturbative Quantum Chromodynamics (QCD). Due to the lack of complete understanding of the non-perturbative QCD so far, one needs to utilize phenomenological models. In describing various processes, such models have achieved relative successes up to now. Among these models the Nambu--Jona-Lasinio (NJL) model~\cite{RevModPhys.64.649,Buballa:2003qv,Cloet:2014rja,Cui:2017ilj,Cui:2016zqp,Zhang:2016zto,Zhang:2018ouu,Zhang:2024dhs,Zhang:2024adr} is a Poincar\'{e} covariant quantum field theory with many of the same low-energy properties as QCD. It has the same flavor symmetries as QCD and should therefore offer a stable chiral effective theory of QCD valid at low to intermediate energies. Above all, it includes the key prominent phenomena of confinement and dynamical chiral symmetry breaking.
	
	The basic degrees of freedom of QCD are quarks and gluons, LFWFs~\cite{Lepage:1980fj,Miller:2009fc,Brodsky:2003pw,Brodsky:2010vs,Ke:2011mu,Ji:2021znw,Zhang:2021shm,Zhang:2021tnr,Lappi:2020ufv} serve as the interpolating functions that connect hadrons with their underlying quark and gluon constituents. LFWF is proved to be Lorentz invariant and expressed in terms of the momentum fractions of the constituents which are independent of the total hadron momentum. As a result, LFWFs play a crucial role in the light-front quantization approach to QCD and our understanding of hadron structure. Therefore, gaining insight into how to derive LFWFs from a fundamental perspective is highly beneficial.
	
	A variety of high energy hadronic processes and experimentally observable quantities can be interpreted in terms of LFWFs, for example, electromagnetic form factors~\cite{Bartel:1973rf,Hohler:1976ax,Vogel:1989iv,Mergell:1995bf,Kelly:2004hm,Perdrisat:2006hj,Puckett:2010ac,DeMori:2022atz,Stamen:2022uqh,Batelaan:2022fdq} and PDFs~\cite{Holt:2010vj,Nguyen:2011jy,Dulat:2015mca,Ding:2019qlr,Cui:2020dlm,Cui:2020tdf} can be analyzed using LFWFs. In addition, LFWFs also can be used to estimate of weak decay rates~\cite{Brodsky:1998hn,Choi:2007yu}, calculate single spin asymmetries in semi-inclusive deep inelastic scattering~\cite{Brodsky:2002cx,Gamberg:2007wm}, and determine the cross sections of electromagnetic production for vector mesons~\cite{Radyushkin:1996ru,Vanderhaeghen:1999xj,Goloskokov:2006hr}. Furthermore, the three-dimensional structure of the nucleon in the momentum and coordinate space characterized by the TMDs~\cite{Collins:2007ph,Meissner:2007rx,Hautmann:2007uw,Barone:2010zz,Puhan:2023hio,Xie:2022lra,Puhan:2023ekt} and generalized parton distributions (GPDs)~\cite{Mueller:1998fv,Ji:1996ek,Radyushkin:1997ki,Vanderhaeghen:1999xj,Belitsky:2001ns,Goeke:2001tz,Burkardt:2002hr,Diehl:2003ny,Belitsky:2005qn,Zhang:2020ecj,Zhang:2021mtn,Zhang:2021uak,Adhikari:2021jrh,Fu:2022bpf,Zhang:2022zim,Luan:2024vgv}. TMDs are sensitive to spin dependent correlations and the parton’s intrinsic transverse motion, offering a more complete understanding of the parton structures, especially the transverse structures of hadrons.
	
	Computing light vector meson LFWFs in QCD presents a significant challenge. The diagonalization of the light-cone QCD Hamiltonian becomes exceedingly challenging as the number of Fock-states~\cite{Brodsky:1997de} increases, due to the numerous possibilities for the creation and annihilation of light-quarks and gluons. By calculating the hadronic eigensolutions $|\psi_H\rangle$ of the QCD light-front Hamiltonian $H_{LF}$ projected on the free Fick basis $\psi_n^H=\langle n|\psi_H\rangle$, one can obtain LFWFs. In principle, it is possible to solve for the hadronic LFWFs directly from fundamental theory by employing nonperturbative techniques such as discretized light-front quantization, the transverse lattice, lattice gauge theory moments, or Bethe-Salpeter (BSE)/Dyson-Schwinger equation (DSE)~\cite{Shi:2021taf,Shi:2022erw}. Reviews of nonperturbative light-front methods can be found in Refs.~\cite{Carbonell:1998rj,Brodsky:1997de,Dalley:2002ug}.
	
	In this paper, we have computed the leading Fock-state LFWFs of the $\rho$ meson within the framework of the NJL model. 
	The NJL model is characterized by an effective Lagrangian for relativistic fermions interacting through local fermion-fermion couplings, and it finds widespread application in various fields.
	In the $\rho$ meson channel, the model is equivalent to the symmetry-preserving treatment of a contact interaction \cite{Gutierrez-Guerrero:2010waf, Roberts:2011wy}; hence, may be seen as an application of continuum Schwinger function methods \cite{Ding:2022ows}.
	Viewed from this perspective, our calculation may be understood as an analysis of $\rho$ meson properties at the hadron scale, $\zeta_{\cal H}$, called $Q_0$ herein, whereat dressed quark and antiquark degrees-of-freedom carry all hadron properties. 
	We do not consider evolution to resolving scales $Q > Q_0$; but that may be accomplished using the all-orders evolution scheme detailed elsewhere \cite{Yin:2023dbw}.
	
	This paper is organized as follows: In Sec.~\ref{nice}, we provide an introduction to the NJL model and then proceed to calculate the $\rho$ meson LFWFs within the framework of the NJL model. In Sec.~\ref{good}, we calculate the TMDs of the $\rho$ meson from the overlap representation of the LFWFs. The $\bm{k}_{\perp}$-weighted moments and the $x$-dependent average transverse momentum $\langle k_{\perp}^n(x)\rangle_{\alpha}$ of $\rho$ TMDs are studied. The positive bounds for $\rho$ meson TMDs and PDFs are checked. Additionally, we examine the spin densities of $\rho$ meson in transverse momentum space. Finally, a concise summary and outlook are provided in Section~\ref{excellent}.

	\section{$\rho$ meson LFWFs in the NJL model}\label{nice}

	\subsection{NJL Model}
	
	The NJL model focuses on quarks as the explicit degrees of freedom, with gluons only expressed implicitly via the infrared strength of the contact interaction. The Lagrangian for SU(2) flavor in the $qq$ channel is given by
	\begin{align}\label{1}
		\mathcal{L}&=\bar{\psi }\left(i\gamma ^{\mu }\partial _{\mu }-\hat{m}\right)\psi\nonumber\\
		&+ G_{\pi }\left[\left(\bar{\psi }\psi\right)^2-\left( \bar{\psi }\gamma _5 \vec{\tau }\psi \right)^2\right]-G_{\omega}\left(\bar{\psi }\gamma _{\mu}\psi\right)^2\nonumber\\
		&-G_{\rho}\left[\left(\bar{\psi }\gamma _{\mu} \vec{\tau } \psi\right)^2+\left( \bar{\psi }\gamma _{\mu}\gamma _5 \vec{\tau } \psi \right)^2\right],
	\end{align}
	where $\psi$ is the quark field, $\hat{m}\equiv\text{diag}\left(m_u,m_d\right)$ is current quark mass matrix and in the isospin symmetry $m_u = m_d =m$. $\vec{\tau}$ are the Pauli matrices represent isospin. In chiral channels the 4-fermion coupling constants are labeled by $G_{\pi}$ , $G_{\omega}$, and $G_{\rho}$.
	In working with Eq.\,\eqref{1}, our calculations deliver properties of what will emerge as the dynamically generated dressed-quark + dressed-antiquark core of the $\rho$ meson.  This is known to be a good approximation -- see, e.g., Ref.\,\cite{Roberts:2011wy}.
	
	In the NJL model, the solution for the dressed quark propagator is,
	\begin{align}\label{2}
		S(k)=\frac{1}{{\not\!k}-M+i \varepsilon}.
	\end{align}

	A regularization scheme is needed to fully define the model because NJL model is not renormalizable. We will implement the proper time regularization (PTR) scheme~\cite{Ebert:1996vx,Hellstern:1997nv,Bentz:2001vc}.
	\begin{align}\label{4}
		\frac{1}{X^n}&=\frac{1}{(n-1)!}\int_0^{\infty}\mathrm{d}\tau\, \tau^{n-1}e^{-\tau X}\nonumber\\
		& \rightarrow \frac{1}{(n-1)!} \int_{1/\Lambda_{\text{UV}}^2}^{1/\Lambda_{\text{IR}}^2}\mathrm{d}\tau\, \tau^{n-1}e^{-\tau X}
	\end{align}
	where $X$ represents a product of propagators that have been combined using Feynman parametrization. The infrared cutoff should be approximately at the scale of $\Lambda_{\text{QCD}}$, and we have selected $\Lambda_{\text{IR}}=0.240$ GeV, as it incorporates the confinement scale into our calculations. The coupling strength $G_{\pi}$, the momentum cutoff $\Lambda_{\text{UV}}$, and the current quark mass $m$ are determined through the Gell-Mann–Oakes–Renner (GMOR) relation, which is expressed as $f_{\pi}^2m_{\pi}^2=-m\langle\bar{\psi}\psi \rangle$. Additionally, the gap equation $M=m-2G_{\pi}\langle\bar{\psi}\psi \rangle$ is utilized in this determination. $\langle\bar{\psi}\psi \rangle$ represents a two-quark condensate derived from QCD sum rules. Here, $m_\pi=0.140$ GeV represents the physical pion mass, while $f_\pi=0.092$ GeV denotes the pion decay constant. $m$ and $M$ refer to the current and constituent quark masses, respectively.
	

	The self-energy of the pseudoscalar bubble diagram is
	\begin{align}\label{ab35}
		&\Pi_{PP}(Q^2)
		=-\frac{3 }{2\pi ^2} \int_0^1\mathrm{d}x\int_{1/\Lambda_{\text{UV}}^2}^{1/\Lambda_{\text{IR}}^2}\mathrm{d}\tau\,\frac{1}{\tau^2}e^{-\tau M^2}\nonumber\\
		&+\frac{3 }{4\pi ^2} \int_0^1\mathrm{d}x\int_{1/\Lambda_{\text{UV}}^2}^{1/\Lambda_{\text{IR}}^2}\mathrm{d}\tau\, \frac{Q^2}{\tau}\,  e^{-\tau(M^2+x\bar{x}Q^2) },
	\end{align}
	where $\bar{x}=1-x$, and $m_{\pi}=0.14$ GeV coincides with the value obtained from the pole condition $1+2G_{\pi} \Pi_{PP} (-m_{\pi}^2)=0$. The values of $G_{\omega}$ and $G_{\rho}$ are determined by $m_{\omega}=0.782$ GeV, $m_{\rho}=0.770$ GeV through the equation $1+2G_i \Pi_{VV} (-m_i^2)=0$, where $i=(\omega,\rho)$. Here, $\Pi_{VV} (Q^2)$ represents the self-energy of the vector bubble diagram as defined in Eq. (\ref{vvbu}). The parameters utilized in this study are listed in Table \ref{tb1}.
	
	The absence of an explicit $Q^2$ scale is a common limitation in most model determinations of quark distributions. As a result, the model scale, denoted as $Q_0^2$, must be established through comparison with empirical data. We adopt $Q_0^2=0.16$ GeV$^2$, consistent with Ref.~\cite{Cloet:2007em}, which is representative of models dominated by valence contributions~\cite{Schreiber:1991tc,Mineo:1999eq,Mineo:2002bg,Ninomiya:2017ggn}.
	
	The form factors of the dressed quark associated with electromagnetic currents are derived from the inhomogeneous Bethe-Salpeter equation~\cite{Zhang:2021shm}
	\begin{align}\label{bsam}
		F_{1i}(Q^2)=\frac{1}{1+2G_i \Pi_{VV}(Q^2)},\quad \quad F_{2i}(Q^2)=0,
	\end{align}
	where $i=(\omega,\rho)$, $\Pi_{VV}$ is the self-energy of the vector bubble diagram
	\begin{align}\label{vvbu}
		\Pi_{VV}(Q^2)=\frac{3 }{\pi ^2} \int_0^1\mathrm{d}x\int_{1/\Lambda_{\text{UV}}^2}^{1/\Lambda_{\text{IR}}^2}\mathrm{d}\tau\, \frac{x \bar{x}}{\tau} Q^2 e^{-\tau(M^2+x\bar{x}Q^2) }.
	\end{align}
	%

	\begin{center}
		\begin{table}
			\caption{Parameters used in our work. The dressed quark mass and regularization parameters are in units of GeV, while coupling constant is in units of GeV$^{-2}$. }\label{tb1}
			\begin{tabular}{p{0.7cm} p{0.8cm} p{0.5cm} p{0.8cm}p{0.8cm}p{0.7cm}p{0.7cm}p{0.7cm}p{0.8cm}p{0.8cm}}
				\hline\hline
				$\Lambda_{\text{IR}}$ & $\Lambda_{\text{UV}}$ & $M$ & $m$ & $m_{\rho}$ & $G_{\pi}$ &$G_{\omega}$&$G_{\rho}$ & $N_T$& $N_L$  \\
				\hline
				0.24 & 0.645 & 0.4 & 0.016 & 0.77 & 19.0 & 10.4 & 11.0 &3.618&3.005 \\
				\hline\hline
			\end{tabular}
		\end{table}
	\end{center}
	
	\subsection{The definition and calculation of LFWFs}\label{qq}
	%
	%
	%
	%
	%
	%
	Considering dynamical spin effects, the vector meson wave functions in the NJL model can be written as
	\begin{align}\label{a93}
		&\Psi_{\lambda ,\lambda '}^{\Lambda}(x,\bm{k}_{\perp})=iN_{\Lambda}\int \frac{\mathrm{d}k^+ \mathrm{d}k^-}{2 \pi }\delta (x-\frac{k^+}{q^+})\nonumber\\
		&\times \frac{\sqrt{x \bar{x}}}{\left(k^2-M^2\right) \left((k-q)^2-M^2\right)}\chi_{\lambda ,\lambda '}^{\Lambda}(x,\bm{k}_{\perp}),
	\end{align}
	where $a^{\pm}=\frac{1}{\sqrt{2}}(a^0\pm a^3)$, $\bm{a}_{\perp}=(a_1,a_2)$ are the light-cone $\pm$ and transverse components of a 4-vector $a^{\mu}$. The Lorentz invariant spin structure for the vector meson is expressed by accounting the photon-quark-antiquark vertex:
	\begin{align}\label{a93}
		\chi_{\lambda ,\lambda '}^{\Lambda}(x,\bm{k}_{\perp})&=\frac{\bar{u}_{\lambda}(xq^+,\bm{k}_{\perp})}{\sqrt{x}}\epsilon_{\Lambda}\cdot \gamma \frac{v_{\lambda'}(\bar{x}q^+,-\bm{k}_{\perp})}{\sqrt{1-x}},
	\end{align}
	where $xq^+$ and $\bar{x}q^+$ are the longitudinal momentum of the quark and antiquark, $\bm{k}_{\perp}$ and $-\bm{k}_{\perp}$ are the transverse momentum of the quark and qntiquark respectively. $\epsilon_{\Lambda}$ are the polarizations vectors, for the longitudinally polarized and the transversely polarized $\rho$ meson are given by
	\begin{align}\label{a93}
		\epsilon_{L}=(\frac{q^+}{m_{\rho}},-\frac{m_{\rho}}{q^+},0,0),\quad \quad \epsilon_T^{\pm}=\mp\frac{1}{\sqrt{2}}(0,0,1,\pm i),
	\end{align}
	then we obtain the final form of the LFWFs
	\begin{align}\label{lf1}
		\Psi_{\lambda ,\lambda '}^{0}&=\frac{\sqrt{6}N_L(x \bar{x}m_{\rho}^2+M^2+\bm{k}_{\perp}^2)\delta _{\lambda ,-\lambda '}}{(\bm{k}_{\bot }^2+M^2-\bar{x} xm_{\rho}^2)} ,
	\end{align}

	\begin{align}\label{lf2}
		\Psi_{\lambda ,\lambda '}^{+1}&=\frac{\sqrt{6}N_T\left(\delta _{\lambda+ ,\lambda'-}xk_{\lambda} - k_{-\lambda}\bar{x}\delta _{\lambda- ,\lambda '+}+M\delta _{\lambda+ ,\lambda'+}\right)}{\bm{k}_{\bot }^2+M^2- \bar{x} xm_{\rho}^2} ,
	\end{align}
	
	\begin{align}\label{lf3}
		\Psi_{\lambda ,\lambda '}^{-1}&=\frac{\sqrt{6}N_T\left(\delta _{\lambda- ,\lambda'+}xk_{\lambda}- k_{-\lambda}\bar{x}\delta _{\lambda +,\lambda '-}-M\delta _{\lambda- ,\lambda'-}\right)}{\bm{k}_{\bot }^2+M^2- \bar{x} xm_{\rho}^2},
	\end{align}
	where $k_{\lambda}=k_i+i \lambda k_2$, $\delta _{ab ,cd}=\delta _{a,b}\delta _{c,d}$, with $\delta _{a,b}$ the Kronecker delta. The LFWFs are divergent, we should apply the PTR to make them convergent. $\Psi_{\pm,\pm}^{\Lambda=0}$ and $\Psi_{\pm,\pm}^{\Lambda=\pm1}$ represent $|l_z|=1$ and $|l_z|=2$ respectively.
	
	The normalization constants $N_{T(L)}$ are determined by
	\begin{align}\label{a93}
		\sum_{\lambda,\lambda'}\int \frac{ \mathrm{d}x\mathrm{d}^2\bm{k}_{\perp}}{2(2\pi)^3} |\Psi_{\lambda ,\lambda '}^{\Lambda}|^2=1,
	\end{align}
	depending upon the polarization of the $\rho$ meson. We obtain $N_T^2=13.092$ and $N_L^2=9.032$.

	\section{$\rho$ meson TMDs}\label{good}
	The transverse momentum-dependent quark correlation function of spin-1 particle is defined as:
	\begin{align}\label{tmdpcf}
		&\bm{\Phi}_{\beta\alpha}^{(\lambda)\bm {S}}(x,\bm{k}_{\perp})\nonumber\\
		&=\int \frac{\mathrm{d}k^+\mathrm{d}k^-}{(2\pi)^4}\delta(k^+-xp^+)\nonumber\\
		&\times \int \mathrm{d}^4z e^{ik\cdot z} \,_{\bm{S}} \langle p,\lambda|\bar{\psi}_{\alpha}(0)\psi_{\beta}(z)|p,\lambda\rangle_{\bm{S}},\nonumber\\
		&=\int \frac{\mathrm{d}z^-\mathrm{d}^2\bm{z}_{\perp}}{(2\pi)^3}\,_{\bm{S}} \langle p,\lambda|\bar{\psi}_{\alpha}(0)\psi_{\beta}(z^-,\bm{z}_{\perp})|p,\lambda\rangle_{\bm{S}}\nonumber\\
		&\times e^{i(xp^+z^--\bm{k}_{\perp}\cdot\bm{z}_{\perp} )}\nonumber\\
		&=\varepsilon_{(\lambda)\mu}^{*}(p)\bm{\Phi}_{\beta\alpha}^{\mu\nu}(x,\bm{k}_{\perp})\varepsilon_{(\lambda)\nu}(p),
	\end{align}
	where $\psi$ stands for the flavor SU(2) quark field operator, $\alpha$ and $\beta$ are Dirac indices, for any four-vector. The 3-momentum of the target ($\bm{p}$) is assumed to be in the $z$-direction, and the quark’s momentum components perpendicular to this direction are denoted by $\bm{k}_{\perp}$. The state $|p,\lambda\rangle$ denotes that the projection of the target's spin on the direction $\bm{S}$ is equal to $\lambda=\pm1$, $0$, and is normalized according to the covariant normalization $\langle p,\lambda^{'}|p,\lambda\rangle=2p^+V\delta_{\lambda,\lambda^{'}}$, where $V$ represents the quantization volume.
	
	At leading-twist, the coefficient functions are represented by
	\begin{align}\label{a93}
		&\langle\gamma^+\rangle_{\bm{S}}^{(\lambda)}(x,\bm{k}_{\perp})=\frac{1}{2}\text{Tr}_D\left[\gamma^+ \Phi^{(\lambda)\bm{S}}(x,\bm{k}_{\perp})\right]\nonumber\\
		&\equiv \varepsilon_{(\lambda)\mu}^{*}(p)\langle\gamma^+\rangle^{\mu\nu}(x,\bm{k}_{\perp})\varepsilon_{(\lambda)\nu}(p),
	\end{align}
	\begin{align}\label{a93}
		&\langle\gamma^+\gamma_5\rangle_{\bm{S}}^{(\lambda)}(x,\bm{k}_{\perp})=\frac{1}{2}\text{Tr}_D\left[\gamma^+\gamma^5\Phi^{(\lambda)\bm{S}}(x,\bm{k}_{\perp})\right]\nonumber\\
		&\equiv \varepsilon_{(\lambda)\mu}^{*}(p)\langle\gamma^+\gamma_5\rangle^{\mu\nu}(x,\bm{k}_{\perp})\varepsilon_{(\lambda)\nu}(p),
	\end{align}
	\begin{align}\label{a93}
		&\langle \gamma^+\gamma^i\gamma^5\rangle_{\bm{S}}^{(\lambda)}(x,\bm{k}_{\perp})=\frac{1}{2}\text{Tr}_D\left[\gamma^+\gamma^i\gamma^5\Phi^{(\lambda)\bm{S}}(x,\bm{k}_{\perp})\right]\nonumber\\
		&\equiv \varepsilon_{(\lambda)\mu}^{*}(p)\langle \gamma^+\gamma^i\gamma^5\rangle^{\mu\nu}(x,\bm{k}_{\perp})\varepsilon_{(\lambda)\nu}(p),
	\end{align}
	where $i = 1, 2$ represents the transverse vector components. According to Refs.~\cite{Bacchetta:2001rb,Ninomiya:2017ggn}, the coefficient functions of the spin-1 hadron TMDs can be expressed as follows: 
	\begin{align}\label{a93}
		&\langle\gamma^+\rangle_{\bm{S}}^{(\lambda)}(x,\bm{k}_{\perp}^2)\equiv f_1(x,\bm{k}_{\perp}^2)+S_{LL}f_{LL}(x,\bm{k}_{\perp}^2)\nonumber\\
		&+\frac{\bm{S}_{LT}\cdot \bm{k}_{\perp}}{m_{\rho}}f_{LT}(x,\bm{k}_{\perp}^2)+\frac{ \bm{k}_{\perp}\cdot \bm{S}_{TT}\cdot \bm{k}_{\perp}}{m_{\rho}^2}f_{TT}(x,\bm{k}_{\perp}^2),
	\end{align}
	\begin{align}\label{a93}
		\langle\gamma^+\gamma^5\rangle_{\bm{S}}^{(\lambda)}(x,\bm{k}_{\perp}^2)&\equiv \lambda \left[S_L g_L(x,\bm{k}_{\perp}^2)+\frac{\bm{S}_{T}\cdot \bm{k}_{\perp}}{m_{\rho}}g_T(x,\bm{k}_{\perp}^2)\right],
	\end{align}
	\begin{align}\label{a93}
		&\langle \gamma^+\gamma^i\gamma^5\rangle_{\bm{S}}^{(\lambda)}(x,\bm{k}_{\perp}^2)\equiv \lambda \left[S_T^i h(x,\bm{k}_{\perp}^2)+S_L \frac{k_{\perp}^i}{m_{\rho}}h_{L}^{\perp}(x,\bm{k}_{\perp}^2)\right]\nonumber\\
		&+ \frac{\lambda}{2m_{\rho}^2}(2k_{\perp}^i \bm{k}_{\perp}\cdot \bm{S}_{T}  -S_T^i\bm{k}_{\perp}^2)h_T^{\perp}(x,\bm{k}_{\perp}^2),
	\end{align}
	where the specific $\bm{S}$ and $\lambda$ dependence are:
	\begin{subequations}
		\begin{align}
			S_{LL}&=(3\lambda^2-2)(\frac{1}{6}-\frac{1}{2}S_L^2)\,,\\
			S_{LT}^i&=(3\lambda^2-2)S_LS_T^i\,,\\
			S_{TT}^{ij}&=(3\lambda^2-2)(S_T^iS_T^j-\frac{1}{2}\bm{S}_T^2\delta^{ij}).
		\end{align}
	\end{subequations}
	$f_{LL}(x,\bm{k}_{\perp}^2)$, $f_{LT}(x,\bm{k}_{\perp}^2)$ and $f_{TT}(x,\bm{k}_{\perp}^2)$ are specific to tensor polarized hadrons with spin $J \geq 1$.
	
	\subsection{Overlap formalism}
	%
	\begin{figure*}
		\centering
		\includegraphics[width=0.35\textwidth]{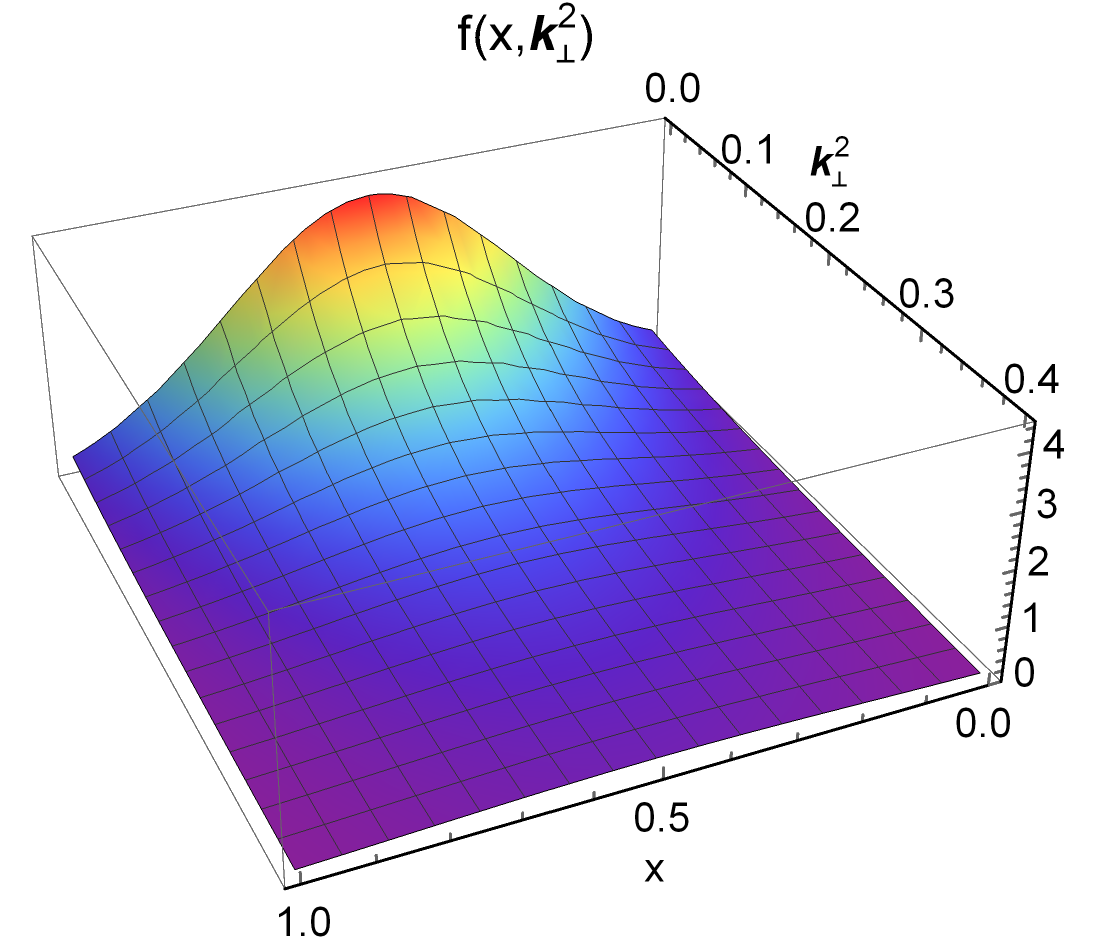}
		\includegraphics[width=0.030\textwidth]{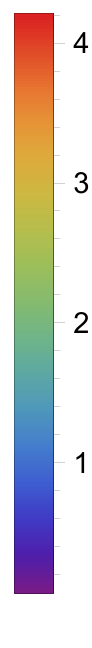}
		\qquad\qquad\qquad
		\includegraphics[width=0.35\textwidth]{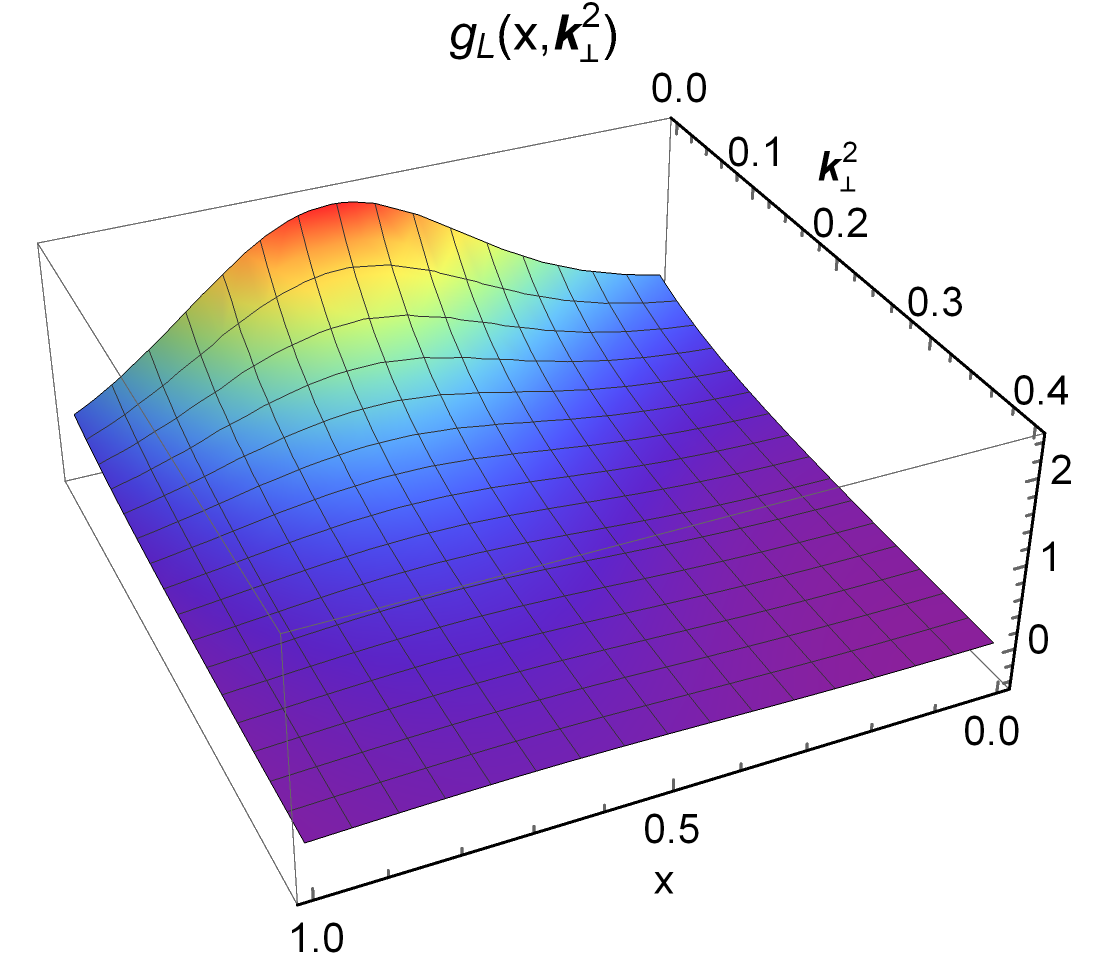}
		\includegraphics[width=0.037\textwidth]{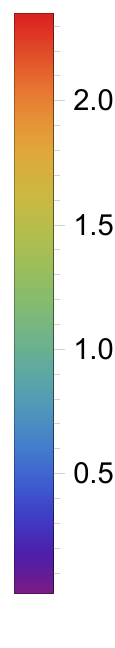}
		\qquad
		\includegraphics[width=0.35\textwidth]{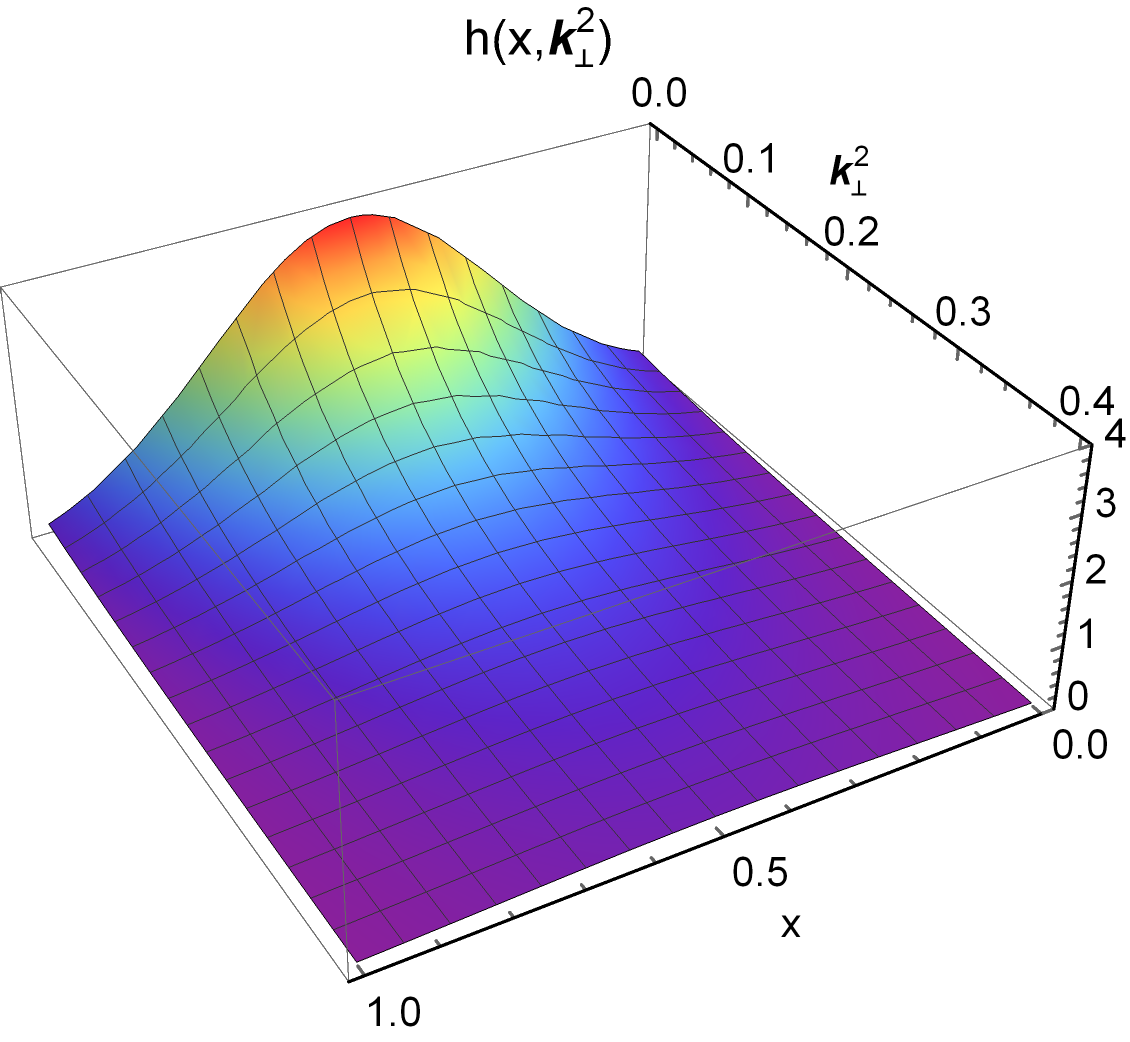}
		\includegraphics[width=0.037\textwidth]{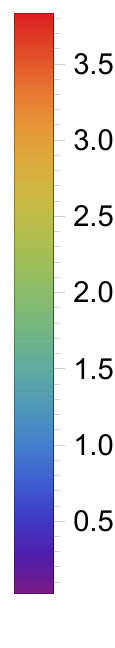}
		\qquad\qquad\qquad
		\includegraphics[width=0.35\textwidth]{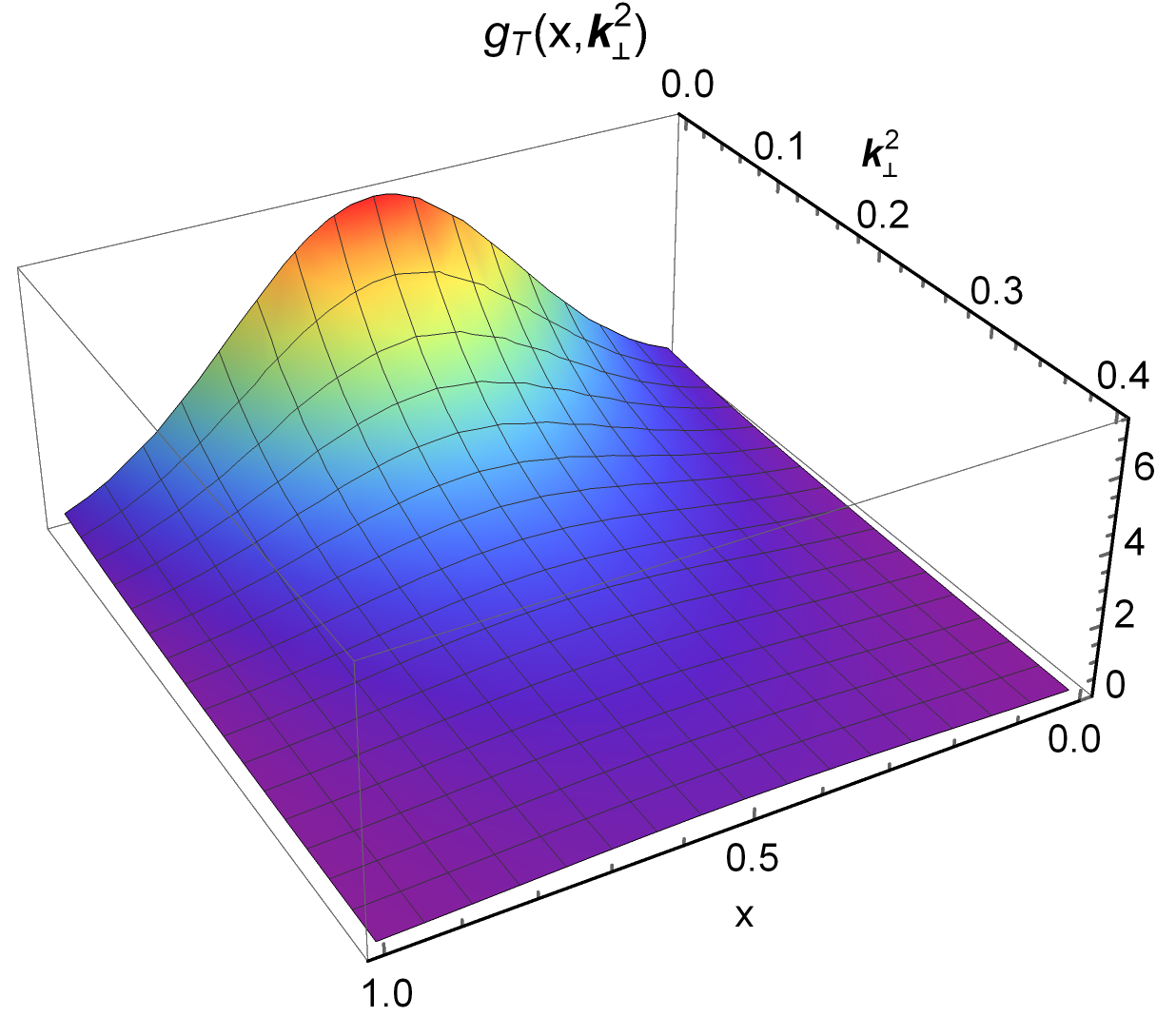}
		\includegraphics[width=0.030\textwidth]{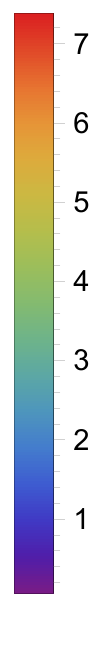}
		\qquad
		\includegraphics[width=0.35\textwidth]{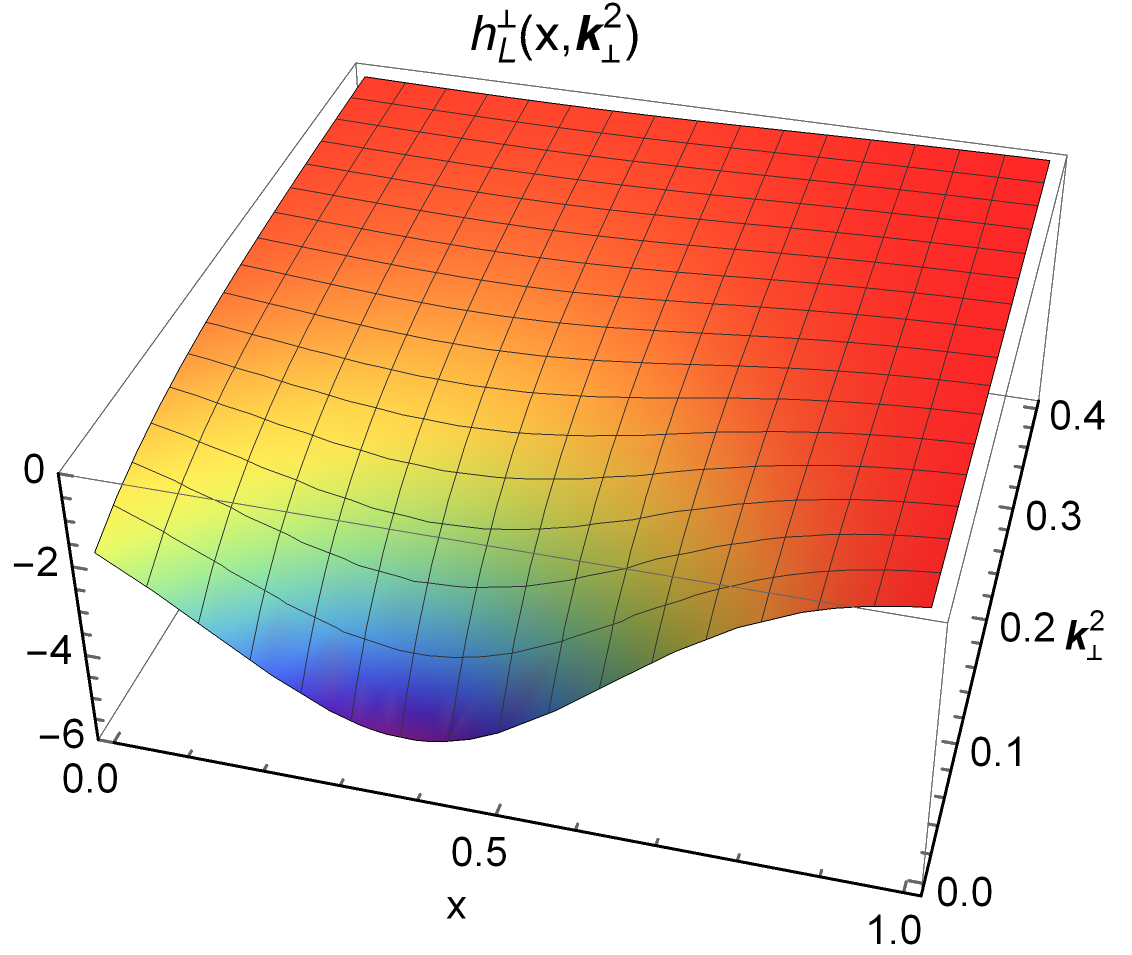}
		\includegraphics[width=0.037\textwidth]{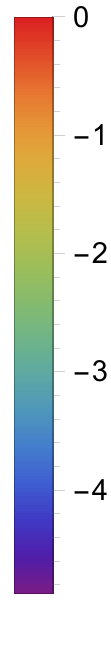}
		\qquad\qquad\qquad
		\includegraphics[width=0.35\textwidth]{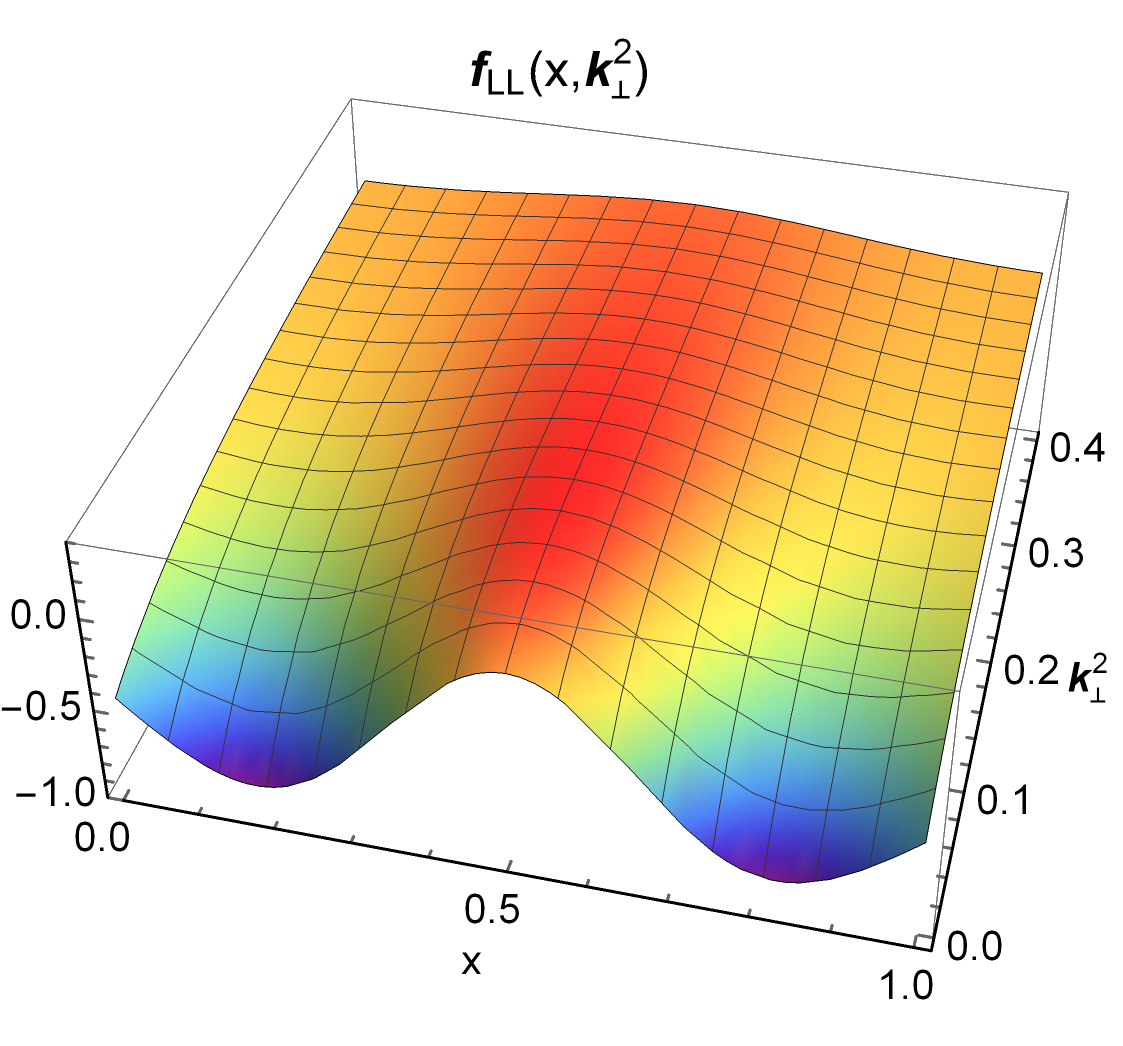}
		\includegraphics[width=0.044\textwidth]{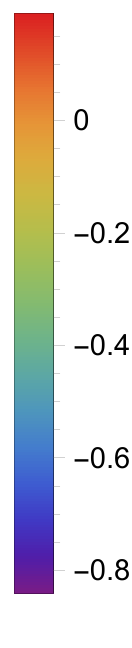}
		\qquad
		\includegraphics[width=0.35\textwidth]{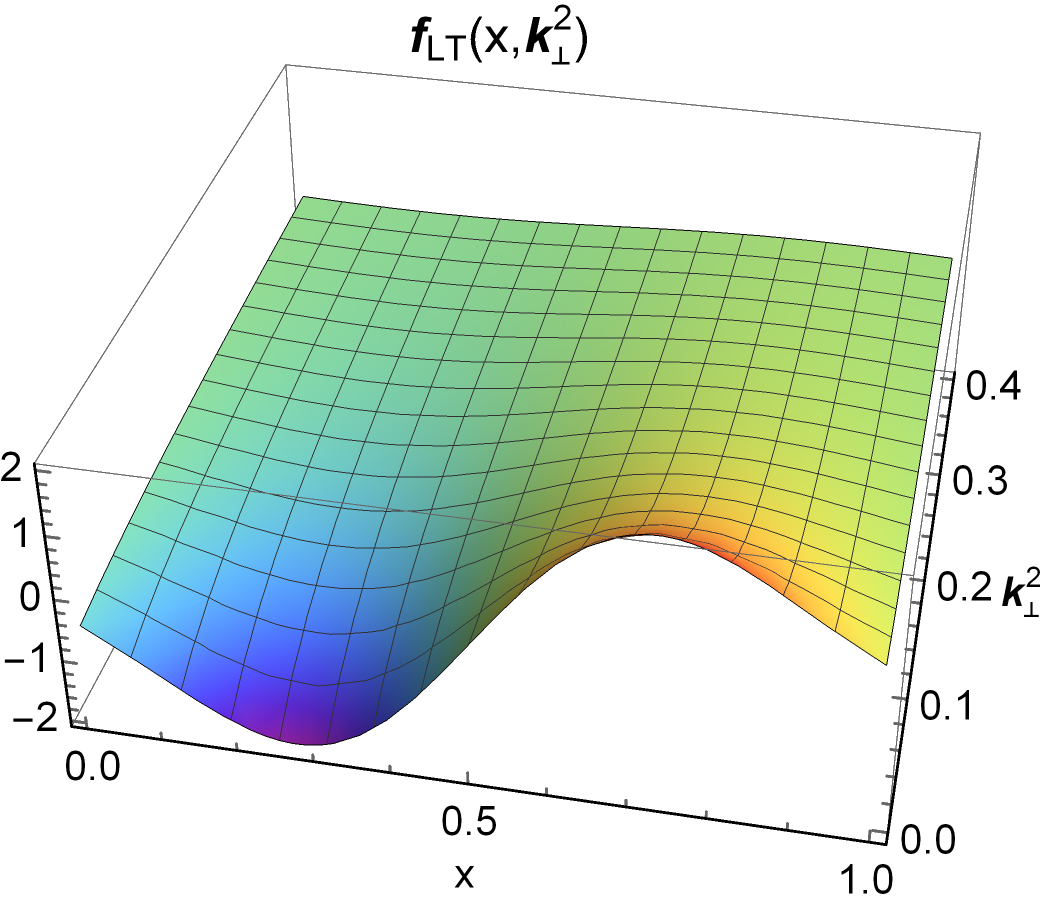}
		\includegraphics[width=0.037\textwidth]{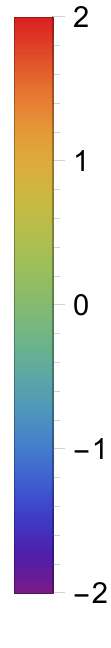}
		\qquad\qquad\qquad
		\includegraphics[width=0.35\textwidth]{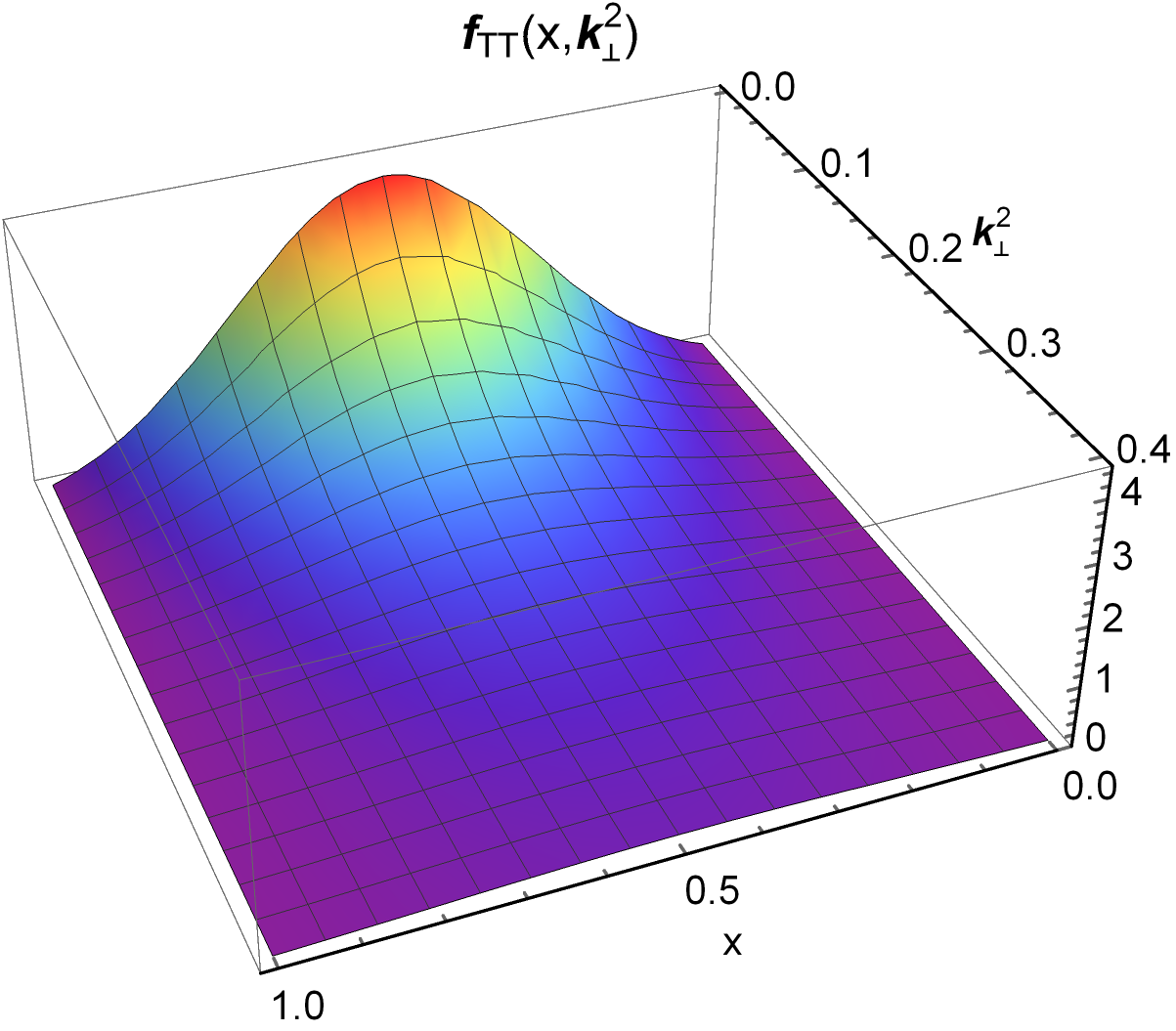}
		\includegraphics[width=0.030\textwidth]{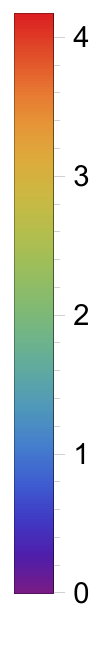}
		\caption{Results for the quark TMDs of $\rho$ meson as a function of $x$ and $\bm{k}_{\perp}^2$ at the model scale $Q_0^2=0.16$ GeV$^2$. The TMDs are given in units of GeV$^{-2}$ and $\bm{k}_{\perp}^2$ in units of GeV$^2$.}\label{tmd1}
	\end{figure*}
	The explicit overlap form for all the T-even TMDs in terms of helicity amplitudes are expressed as~\cite{Shi:2022erw}
	\begin{widetext}
		\begin{subequations}
			\begin{align}\label{a93}
				f(x,\bm{k}_{\perp}^2)&=\frac{1}{6(2\pi)^3}\sum_{h_q,h_{\bar{q}}}(|\Psi_{h_q,h_{\bar{q}}}^{0}|^2+|\Psi_{h_q,h_{\bar{q}}}^{+1}|^2+|\Psi_{h_q,h_{\bar{q}}}^{-1}|^2)\,, \\
				g_L(x,\bm{k}_{\perp}^2)&=\frac{1}{4(2\pi)^3}\sum_{h_{\bar{q}}}(|\Psi_{+,h_{\bar{q}}}^{+1}|^2-|\Psi_{-,h_{\bar{q}}}^{+1}|^2-|\Psi_{+,h_{\bar{q}}}^{-1}|^2+|\Psi_{-,h_{\bar{q}}}^{-1}|^2)\,, \\
				g_T(x,\bm{k}_{\perp}^2)&=\frac{m_{\rho}}{4\sqrt{2}(2\pi)^3\bm{k}_{\perp}^2}\sum_{h_{\bar{q}}}(k_R(\Psi_{+,h_{\bar{q}}}^{+1*}\Psi_{+,h_{\bar{q}}}^{0}-\Psi_{-,h_{\bar{q}}}^{+1*}\Psi_{-,h_{\bar{q}}}^{0}+\Psi_{+,h_{\bar{q}}}^{0*}\Psi_{+,h_{\bar{q}}}^{-1}-\Psi_{-,h_{\bar{q}}}^{0*}\Psi_{-,h_{\bar{q}}}^{-1}))\nonumber\\
				&+\frac{m_{\rho}}{4\sqrt{2}(2\pi)^3\bm{k}_{\perp}^2}\sum_{h_{\bar{q}}}(k_L(\Psi_{+,h_{\bar{q}}}^{0*}\Psi_{+,h_{\bar{q}}}^{+1}-\Psi_{-,h_{\bar{q}}}^{0*}\Psi_{-,h_{\bar{q}}}^{+1}+\Psi_{+,h_{\bar{q}}}^{-1*}\Psi_{+,h_{\bar{q}}}^{0}-\Psi_{-,h_{\bar{q}}}^{-1*}\Psi_{-,h_{\bar{q}}}^{0}))\,, \\
				h(x,\bm{k}_{\perp}^2)&=\frac{1}{4\sqrt{2}(2\pi)^3}\sum_{h_{\bar{q}}}(\Psi_{+,h_{\bar{q}}}^{+1*}\Psi_{-,h_{\bar{q}}}^{0}+\Psi_{-,h_{\bar{q}}}^{0*}\Psi_{+,h_{\bar{q}}}^{+1}+\Psi_{+,h_{\bar{q}}}^{0*}\Psi_{-,h_{\bar{q}}}^{-1}+\Psi_{-,h_{\bar{q}}}^{-1*}\Psi_{+,h_{\bar{q}}}^{0})\,, \\
				h_L^{\perp}(x,\bm{k}_{\perp}^2)&=\frac{m_{\rho}}{4(2\pi)^3\bm{k}_{\perp}^2}\sum_{h_{\bar{q}}}(k_R(\Psi_{-,h_{\bar{q}}}^{+1*}\Psi_{+,h_{\bar{q}}}^{+1}-\Psi_{-,h_{\bar{q}}}^{-1*}\Psi_{+,h_{\bar{q}}}^{-1})+k_L(\Psi_{+,h_{\bar{q}}}^{+1*}\Psi_{-,h_{\bar{q}}}^{+1}-\Psi_{+,h_{\bar{q}}}^{-1*}\Psi_{-,h_{\bar{q}}}^{-1}))\,, \\
				h_T^{\perp}(x,\bm{k}_{\perp}^2)&=\frac{m_{\rho}^2}{2\sqrt{2}(2\pi)^3\bm{k}_{\perp}^4}\sum_{h_{\bar{q}}}(k_R^2(\Psi_{-,h_{\bar{q}}}^{+1*}\Psi_{+,h_{\bar{q}}}^{0}+\Psi_{-,h_{\bar{q}}}^{0*}\Psi_{+,h_{\bar{q}}}^{-1})+k_L^2(\Psi_{+,h_{\bar{q}}}^{0*}\Psi_{-,h_{\bar{q}}}^{+1}+\Psi_{+,h_{\bar{q}}}^{-1*}\Psi_{-,h_{\bar{q}}}^{0}))\,, \\
				f_{LL}(x,\bm{k}_{\perp}^2)&=\frac{1}{2(2\pi)^3}\sum_{h_q,h_{\bar{q}}}(|\Psi_{h_q,h_{\bar{q}}}^{0}|^2-\frac{1}{2}(|\Psi_{h_q,h_{\bar{q}}}^{+1}|^2+|\Psi_{h_q,h_{\bar{q}}}^{-1}|^2))\,, \\
				f_{LT}(x,\bm{k}_{\perp}^2)&=\frac{m_{\rho}}{4\sqrt{2}(2\pi)^3\bm{k}_{\perp}^2}\sum_{h_{\bar{q}}}(k_R(\Psi_{+,h_{\bar{q}}}^{+1*}\Psi_{+,h_{\bar{q}}}^{0}+\Psi_{-,h_{\bar{q}}}^{+1*}\Psi_{-,h_{\bar{q}}}^{0}-\Psi_{+,h_{\bar{q}}}^{0*}\Psi_{+,h_{\bar{q}}}^{-1}-\Psi_{-,h_{\bar{q}}}^{0*}\Psi_{-,h_{\bar{q}}}^{-1}))\nonumber\\
				&+\frac{m_{\rho}}{4\sqrt{2}(2\pi)^3\bm{k}_{\perp}^2}\sum_{h_{\bar{q}}}(k_L(\Psi_{+,h_{\bar{q}}}^{0*}\Psi_{+,h_{\bar{q}}}^{+1}+\Psi_{-,h_{\bar{q}}}^{0*}\Psi_{-,h_{\bar{q}}}^{+1}-\Psi_{+,h_{\bar{q}}}^{-1*}\Psi_{+,h_{\bar{q}}}^{0}-\Psi_{-,h_{\bar{q}}}^{-1*}\Psi_{-,h_{\bar{q}}}^{0}))\,, \\
				f_{TT}(x,\bm{k}_{\perp}^2)&=\frac{m_{\rho}^2}{4(2\pi)^3\bm{k}_{\perp}^4}\sum_{h_{\bar{q}}}(k_R^2(\Psi_{+,h_{\bar{q}}}^{+1*}\Psi_{+,h_{\bar{q}}}^{-1}+\Psi_{-,h_{\bar{q}}}^{+1*}\Psi_{-,h_{\bar{q}}}^{-1})+k_L^2(\Psi_{+,h_{\bar{q}}}^{-1*}\Psi_{+,h_{\bar{q}}}^{+1}+\Psi_{-,h_{\bar{q}}}^{-1*}\Psi_{-,h_{\bar{q}}}^{+1}))\,,
			\end{align}
		\end{subequations}
	\end{widetext}
	where $k_{R(L)}=k_1\pm ik_2$, using the LFWFs in Eqs. (\ref{lf1}) - (\ref{lf3}), we can derive explicit expressions for the leading-twist T-even TMDs for the $\rho$ meson
	\begin{subequations}\label{tmdf1}
		\begin{align}\label{pc9}
			f(x,\bm{k}_{\perp}^2)&=\frac{2N_L^2}{(2\pi)^3}\frac{\left(x\bar{x}m_{\rho}^2+M^2+\bm{k}_{\perp}^2\right)^2}{[A]^2}\nonumber\\
			&+\frac{2N_T^2}{(2\pi)^3}\frac{(M^2+x^2\bm{k}_{\bot }^2+\bar{x}^2\bm{k}_{\bot }^2)}{[A]^2}\,, \\
			g_L(x,\bm{k}_{\perp}^2)&=\frac{3N_T^2}{(2\pi)^3}\frac{(M^2+(2x-1)\bm{k}_{\bot }^2)}{[A]^2}\,, \\
			g_T(x,\bm{k}_{\perp}^2)&=\frac{6N_TN_Lm_{\rho}}{\sqrt{2}(2\pi)^3}\frac{ \left(x\bar{x}m_{\rho}^2+M^2+\bm{k}_{\perp}^2\right)}{[A]^2}\,, \\
			h(x,\bm{k}_{\perp}^2)&=\frac{6N_TN_L M}{\sqrt{2}(2\pi)^3}\frac{\left(x\bar{x}m_{\rho}^2+M^2+\bm{k}_{\perp}^2\right)}{[A]^2}\,, \\
			h_L^{\perp}(x,\bm{k}_{\perp}^2)&=-\frac{6N_T^2}{(2\pi)^3}\frac{\bar{x}Mm_{\rho}}{[A]^2}\,, \\
			h_T^{\perp}(x,\bm{k}_{\perp}^2)&=0\,, \\
			f_{LL}(x,\bm{k}_{\perp}^2)&=\frac{6N_L^2}{(2\pi)^3}\frac{\left(x \bar{x}m_{\rho}^2+M^2+\bm{k}_{\perp}^2\right)^2}{[A]^2}\nonumber\\
			&-\frac{3N_T^2}{(2\pi)^3}\frac{(M^2+x^2\bm{k}_{\bot }^2+\bar{x}^2\bm{k}_{\bot }^2)}{[A]^2}\,, \\
			f_{LT}(x,\bm{k}_{\perp}^2)&=\frac{6N_TN_Lm_{\rho}}{\sqrt{2}(2\pi)^3}\frac{(2x-1)\left(x\bar{x}m_{\rho}^2+M^2+\bm{k}_{\perp}^2\right)}{[A]^2}\,, \\
			f_{TT}(x,\bm{k}_{\perp}^2)&=\frac{6N_T^2m_{\rho}^2}{(2\pi)^3}\frac{ x \bar{x}}{[A]^2},
		\end{align}
	\end{subequations}
	where $A=\bm{k}_{\perp}^2+M^2+x(x-1)m_{\rho}^2$, these TMDs are divergent, apply the PTR scheme one can obtain the final form of $\rho$ meson TMDs,
	\begin{subequations}\label{tmdj}
		\begin{align}
			f&= \frac{N_L^2}{4\pi^3}\int \mathrm{d}\tau (3x \bar{x}m_{\rho}^2+M^2+\frac{1}{\tau}+ 4 \tau x^2 \bar{x}^2m_{\rho}^4)e^{-\tau A}\nonumber\\
			&+\frac{N_T^2}{4\pi^3}\int \mathrm{d}\tau (x^2+\bar{x}^2)(1+\tau x\bar{x} m_{\rho}^2)e^{-\tau A}\nonumber\\
			&+\frac{N_T^2}{4\pi^3}\int \mathrm{d}\tau \tau(1+2x\bar{x}M^2)e^{-\tau A}\,, \\
			g_L&=\frac{3N_T^2}{8\pi^3}\int \mathrm{d}\tau e^{-\tau A}\nonumber\\
			&\times ((2x-1)(1-\tau(M^2-m_{\rho}^2\bar{x} x))+ \tau M^2)\,, \\
			g_T&=\frac{3N_TN_L}{4\sqrt{2}\pi^3}\int \mathrm{d}\tau m_{\rho}(1+ 2\tau x \bar{x}m_{\rho}^2)e^{-\tau A}\,, \\
			h&=\frac{3N_TN_LM }{4\sqrt{2}\pi^3}\int \mathrm{d}\tau (1+2\tau x \bar{x}m_{\rho}^2)e^{-\tau A}\,, \\
			h_L^{\perp}&=-\frac{3N_T^2}{4\pi^3}\int \mathrm{d}\tau \tau \bar{x}M m_{\rho}e^{-\tau A}\,, \\
			h_T^{\perp}&=0\,, \\
			f_{LL}&=\frac{3N_L^2}{4\pi^3}\int \mathrm{d}\tau  (3x \bar{x}m_{\rho}^2+M^2+\frac{1}{\tau}+ 4\tau x^2 \bar{x}^2m_{\rho}^4)e^{-\tau A}\nonumber\\
			&-\frac{3N_T^2}{8\pi^3}\int \mathrm{d}\tau (x^2+\bar{x}^2)(1+\tau x\bar{x} m_{\rho}^2)e^{-\tau A}\nonumber\\
			&-\frac{3N_T^2}{8\pi^3}\int \mathrm{d}\tau \tau(1+2x\bar{x} M^2)e^{-\tau A}\,, \\
			f_{LT}&=\frac{3N_TN_Lm_{\rho}}{4\sqrt{2}\pi^3}\int \mathrm{d}\tau (2x-1)(1+2\tau x\bar{x}m_{\rho}^2)e^{-\tau A} \,, \\
			f_{TT}&=\frac{3N_T^2}{4\pi^3}\int \mathrm{d}\tau \tau x \bar{x}m_{\rho}^2 e^{-\tau A}.
		\end{align}
	\end{subequations}
	
	The three-dimensional structure of the eight non-zero valance quark TMDs in the NJL model is depicted in Fig. \ref{tmd1}. From the diagram, , it can be observed that the TMDs exhibit symmetry around $x=1/2$, with the exception of $h_{L}^{\perp}(x,\bm{k}_{\perp}^2)$ and $g_L(x,\bm{k}_{\perp}^2)$. For $h_{L}^{\perp}(x,\bm{k}_{\perp}^2)$ the minimum value occurs at approximately $x\simeq0.4$ at $\bm{k}_{\perp}^2\simeq0$, which is smaller than $0.5$, consistent with Refs.~\cite{Kaur:2020emh,Ninomiya:2017ggn}. As $\bm{k}_{\perp}^2$ increases, the minimum value shifts to smaller values of $x$. Regarding $g_L(x,\bm{k}_{\perp}^2)$, the maximum value is around $x\simeq0.6$ for $\bm{k}_{\perp}^2\simeq0$, where $x>1/2$. In Ref.~\cite{Ninomiya:2017ggn}, they also calculated the $\rho$ meson TMDs in the NJL model using a covariant approach by directly calculating them from Feynman diagrams in the NJL model. Our method involves first calculating the LFWFs of $\rho$ meson and then obtaining its TMDs from their overlap. The results of TMDs are similar except for $g_L(x,\bm{k}_{\perp}^2)$. Our calculation yields positive values for all regions of $g_L(x,\bm{k}_{\perp}^2)$, whereas in Ref.~\cite{Ninomiya:2017ggn}, $g_L(x,\bm{k}_{\perp}^2)$ has negative values. This can also be observed in the PDFs, and we will discuss this later.
	
	Similar to Ref.~\cite{Ninomiya:2017ggn}, our TMDs results in Eq. (\ref{tmdj}) show that
	\begin{align}\label{1ts}
		g_T(x,\bm{k}_{\perp}^2)=\frac{m_{\rho}}{M}h(x,\bm{k}_{\perp}^2),
	\end{align}
	this relationship would not maintain QCD evolution. $h(x,\bm{k}_{\perp}^2)$ and $h_L^{\perp}(x,\bm{k}_{\perp}^2)$ are proportional to dressed quark mass $M$.

	\subsection{Transverse momentum dependence}
\begin{figure}
		\centering
		\includegraphics[width=0.5\textwidth]{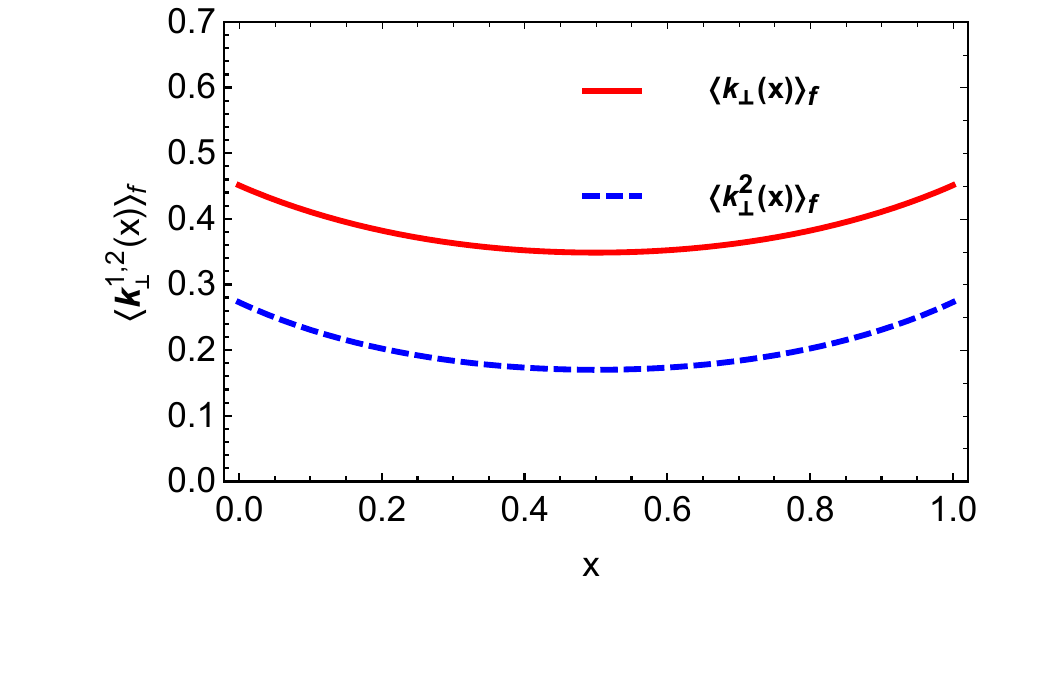}
		\caption{The $x$-dependent average transverse momentum and square: $\langle k_{\perp}(x)\rangle_{f}$ -- solid red curve; $\langle k_{\perp}^2(x)\rangle_{f}$ -- dashed blue curve. }\label{xdatm}
	\end{figure}

	The most intriguing aspect of the TMDs is, undoubtedly, their dependence on transverse momentum. Our initial investigation focuses on the typical transverse momenta of unpolarized quarks in the NJL model. This dependence on transverse momentum of the TMDs can be further elucidated through $\bm{k}_{\perp}$-weighted moments.
	\begin{align}\label{ts}
		\langle k_{\perp}^n\rangle_{\alpha}&\equiv \frac{\int_0^1 \mathrm{d}x\int \mathrm{d}^2\bm{k}_{\perp} |\bm{k}_{\perp}|^n \alpha(x,\bm{k}_{\perp}^2) }{\int_0^1 \mathrm{d}x\int \mathrm{d}^2\bm{k}_{\perp} \alpha(x,\bm{k}_{\perp}^2)},
	\end{align}
	where $n$ represents the order of the moment, $\alpha(x,\bm{k}_{\perp}^2)$ is an arbitrary TMD. The results of $\langle k_{\perp}^n\rangle$ are listed in Table \ref{tb2}, where we also include the NJL results from Ref.~\cite{Ninomiya:2017ggn}. It can be observed from Table \ref{tb2} that the numerical results are similar, except for $g_L$, as our results are larger than theirs. The $\bm{k}_{\perp}$-weighted moments of $g_L$ coincide with those of $h_L^{\perp}$, which is consistent with the findings in Refs.~\cite{Kaur:2020emh,Shi:2022erw}. In our results, both $f_{LL}(x)$ and $f_{LT}(x)$ are zero, leading to a vanishing denominator in the above equation.
	\begin{center}
		\begin{table}
			\caption{The $\langle k_{\perp}\rangle$, $\langle k_{\perp}^2\rangle$ and $\langle k_{\perp}^3\rangle$ moments of the various TMDs defined by Eq. (\ref{ts}), in units of GeV,  GeV$^2$ and GeV$^3$ respectively. Note, the denominator of Eq. (\ref{ts}) vanishes for $f_{LL}(x,\bm{k}_{\perp}^2)$ and $f_{LT}(x,\bm{k}_{\perp}^2)$, so they are not listed. }\label{tb2}
			\begin{tabular}{p{1.2cm}p{0.9cm} p{0.9cm} p{0.9cm} p{0.9cm}p{0.9cm}p{0.9cm}p{0.9cm}}
				\hline\hline
				&$\langle k_{\perp}^n\rangle$&$f$&$g_L$&$g_T$&$h$&$h_L^{\perp}$ &$f_{TT}$\\
				\hline
				Ref.~\cite{Ninomiya:2017ggn}&$\langle k_{\perp}\rangle$&0.32&0.08&0.34&0.34&0.33&0.32\\
				\hline
				Ref.~\cite{Ninomiya:2017ggn}&$\langle k_{\perp}^2\rangle$&0.13&--0.11&0.16&0.16&0.15&0.14\\
				\hline
				Our&$\langle k_{\perp}\rangle$&0.368&0.327&0.346&0.346&0.327&0.322\\
				\hline
				Our&$\langle k_{\perp}^2\rangle$&0.189&0.147&0.165&0.165&0.147&0.143\\
				\hline
				Our&$\langle k_{\perp}^3\rangle$&0.124&0.084&0.100&0.100&0.084&0.080\\
				\hline\hline
			\end{tabular}
		\end{table}
	\end{center}
	In order to obtain the typical transverse momenta of unpolarized quarks in the NJL model, we also introduce an $x$-dependent average transverse momentum $\langle k_{\perp}^n(x)\rangle$, as described in Ref.~\cite{Avakian:2010br}
	\begin{align}\label{ts1}
		\langle k_{\perp}^n(x)\rangle_{\alpha}&= \frac{\int \mathrm{d}^2\bm{k}_{\perp} |\bm{k}_{\perp}|^n \alpha(x,\bm{k}_{\perp}^2) }{\int \mathrm{d}^2\bm{k}_{\perp} \alpha(x,\bm{k}_{\perp}^2)},
	\end{align}
	where $\alpha$ is quark TMDs. The $x$-dependent average transverse momentum $\langle k_{\perp}(x)\rangle_{f}$ and square $\langle k_{\perp}^2(x)\rangle_{f}$ are plotted in Fig. \ref{xdatm}.
	We find that in the valence-$x$ region $\langle k_{\perp}(x)\rangle_{f}$ and $\langle k_{\perp}^2(x)\rangle_{f}$ depend on $x$ weakly. It has a minimum value around $x = 1/2$ and symmetric around $x = 1/2$, the maximum values are around $x=0$ and $x=1$. Numerically we find
	\begin{align}\label{ts2}
		\langle k_{\perp}(x)\rangle_{f}&\approx 0.4 \quad \text{GeV} \,,\\
		\langle k_{\perp}^2(x)\rangle_{f}&\approx 0.25 \quad \text{GeV}^2 ,
	\end{align}
	this weak dependence is similar to the findings in Ref.~\cite{Avakian:2010br} regarding the $u$ quark TMDs of the nucleon, but the numerical values are larger than their results at low hadronic scales. At higher scales, larger values are required as indicated by previous studies~\cite{DAlesio:2004eso,Anselmino:2005nn,Collins:2005rq}, such as $0.64$ GeV from the EMC data in Ref.~\cite{Anselmino:2005nn} and $0.56$ GeV from HERMES data in Ref.~\cite{Collins:2005rq} for the $u$ quark's $x$-dependent average transverse momentum $\langle k_{\perp}(x)\rangle_{f}$ of the nucleon.
	\begin{figure}
		\centering
		\includegraphics[width=0.5\textwidth]{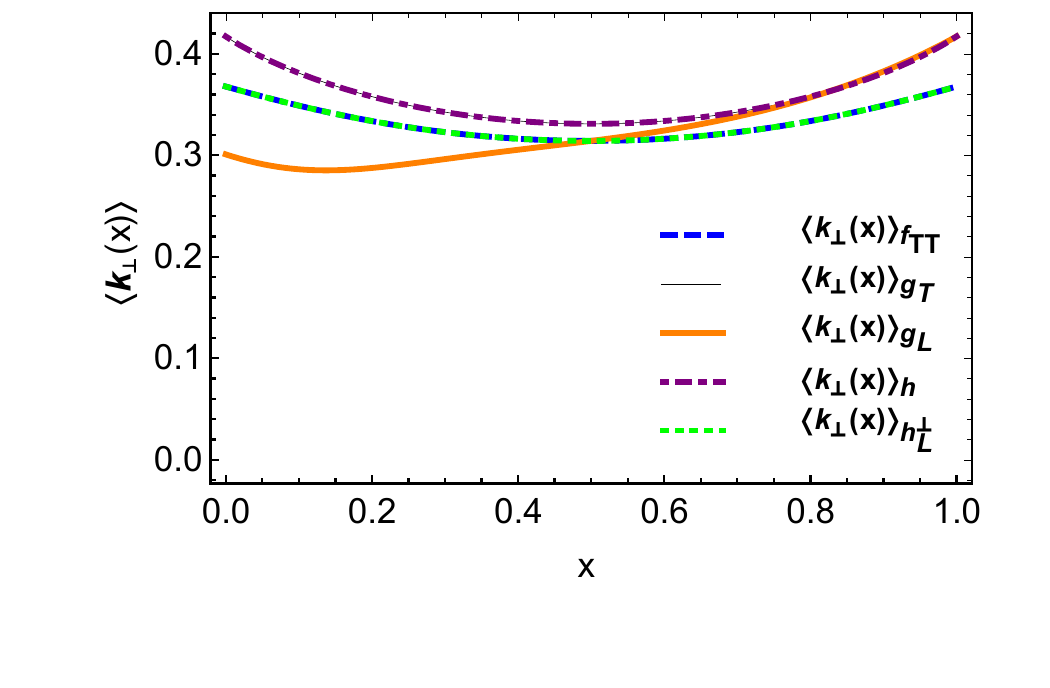}
		\qquad
		\includegraphics[width=0.5\textwidth]{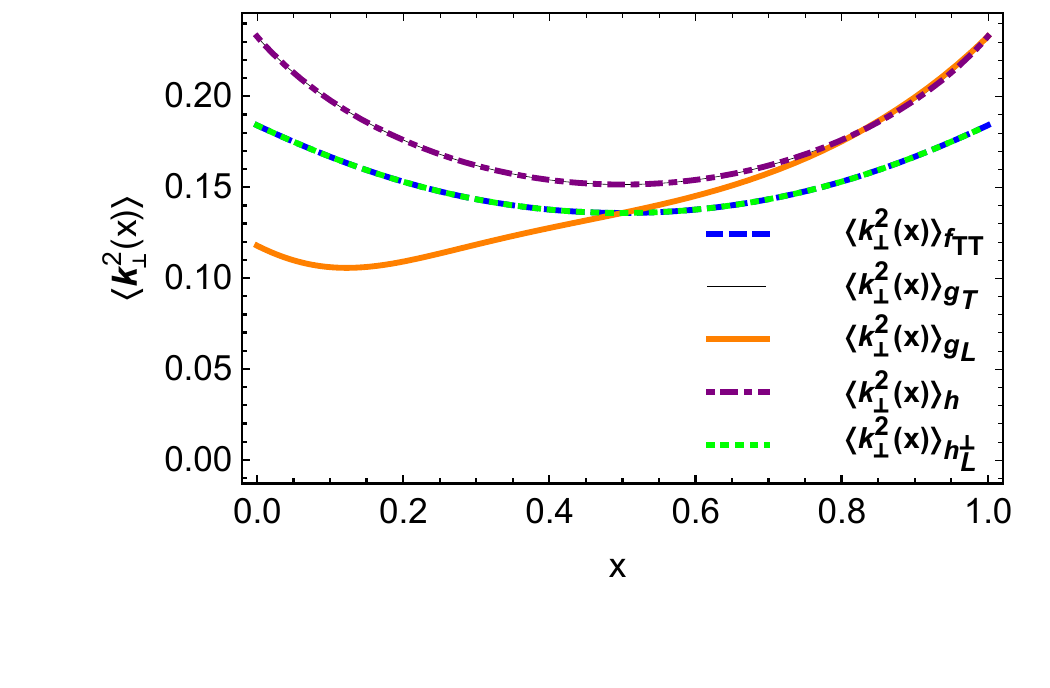}
		\caption{Upper panel: the $x$-dependent average transverse momentum $\langle k_{\perp}(x)\rangle_{\alpha}$. $\langle k_{\perp}(x)\rangle_{f_{TT}}$ -- the dashed blue curve; $\langle k_{\perp}(x)\rangle_{g_T}$ -- the thin solid black curve; $\langle k_{\perp}(x)\rangle_{g_L}$ -- the thick solid orange curve; $\langle k_{\perp}(x)\rangle_{h}$ -- the dot-dashed purple curve; $\langle k_{\perp}(x)\rangle_{h_L^{\perp}}$ -- the dotted green curve; and Lower panel: the $x$-dependent average transverse momentum square $\langle k_{\perp}^2(x)\rangle_{\alpha}$. $\langle k_{\perp}^2(x)\rangle_{f_{TT}}$ -- the dashed blue curve; $\langle k_{\perp}^2(x)\rangle_{g_T}$ -- the thin solid black curve; $\langle k_{\perp}^2(x)\rangle_{g_L}$ -- the thick solid orange curve; $\langle k_{\perp}^2(x)\rangle_{h}$ -- the dot-dashed purple curve; $\langle k_{\perp}^2(x)\rangle_{h_L^{\perp}}$ -- the dotted green curve.}\label{xdatm1}
	\end{figure}

	The $x$-dependent average transverse momentum $\langle k_{\perp}(x)\rangle_{\alpha}$ and square $\langle k_{\perp}^2(x)\rangle_{\alpha}$ of the remanent TMDs are depicted in Fig. \ref{xdatm1}. The plots illustrate that $\langle k_{\perp}^{1,2}(x)\rangle_{g_T}$ and $\langle k_{\perp}^{1,2}(x)\rangle_{h}$ coincide, and they are numerically the largest. Numerically, it is found that
	\begin{align}\label{ts3}
		\langle k_{\perp}(x)\rangle_{g_T}&=\langle k_{\perp}(x)\rangle_{h}\simeq [0.35,0.42] \quad \text{GeV} \,,\\
		\langle k_{\perp}^2(x)\rangle_{g_T}&=\langle k_{\perp}^2(x)\rangle_{h}\simeq [0.16,0.24] \quad \text{GeV}^2 ,
	\end{align}
	$\langle k_{\perp}^{1,2}(x)\rangle_{f_{TT}}$ and $\langle k_{\perp}^{1,2}(x)\rangle_{h_L^{\perp}}$ are also coincide with each other. Their behaviors are also similar, exhibiting a minimum value around $x = 1/2$, symmetry around $x = 1/2$, and maximum values at approximately $x=0$ and $x=1$. This similarity is further supported by Table \ref{tb2}, where it is shown that $\langle k_{\perp}^n\rangle_{g_T}=\langle k_{\perp}^n\rangle_{h}$. Numerically we find
	\begin{align}\label{ts3}
		\langle k_{\perp}(x)\rangle_{f_{TT}}&=\langle k_{\perp}(x)\rangle_{h_L^{\perp}}\simeq [0.32,0.38] \quad \text{GeV} \,,\\
		\langle k_{\perp}^2(x)\rangle_{f_{TT}}&=\langle k_{\perp}^2(x)\rangle_{h_L^{\perp}}\simeq [0.14,0.19] \quad \text{GeV}^2 .
	\end{align}
	Table \ref{tb2} demonstrates that $\langle k_{\perp}^n\rangle_{f_{TT}}$ is not equal to $\langle k_{\perp}^n\rangle_{h_L^{\perp}}$. However, based on Fig. \ref{xdatm1}, we should have 
	%
	\begin{align}\label{ts3}
		\int_0^1 \mathrm{d}x \langle k_{\perp}^n(x)\rangle_{f_{TT}}= \int_0^1 \mathrm{d}x \langle k_{\perp}^n(x)\rangle_{h_L^{\perp}}=0.332,
	\end{align}
	which is slightly larger than $\langle k_{\perp}^n\rangle_{f_{TT}}$ and $\langle k_{\perp}^n\rangle_{h_L^{\perp}}$, this indicates that
	\begin{align}\label{ts3}
		\int_0^1 \mathrm{d}x \langle k_{\perp}^n(x)\rangle_{f_{TT}}&\neq \frac{\int_0^1 \mathrm{d}x \int \mathrm{d}^2\bm{k}_{\perp} |\bm{k}_{\perp}|^n f_{TT} (x,\bm{k}_{\perp}^2) }{\int_0^1 \mathrm{d}x \int \mathrm{d}^2\bm{k}_{\perp} f_{TT}(x,\bm{k}_{\perp}^2)}\,,\\
		\int_0^1 \mathrm{d}x \langle k_{\perp}^n(x)\rangle_{h_L^{\perp}}&\neq \frac{\int_0^1 \mathrm{d}x \int \mathrm{d}^2\bm{k}_{\perp} |\bm{k}_{\perp}|^n h_L^{\perp} (x,\bm{k}_{\perp}^2) }{\int_0^1 \mathrm{d}x \int \mathrm{d}^2\bm{k}_{\perp} h_L^{\perp}(x,\bm{k}_{\perp}^2)}\,,
	\end{align}
	$\langle k_{\perp}^{1,2}(x)\rangle_{g_L}$ is different, the minimum value is around $x =0.1$, the maximum values is around $x=1$, numerically,
	\begin{align}\label{ts3}
		\langle k_{\perp}(x)\rangle_{g_L}&\in [0.28,0.42] \quad \text{GeV} \,,\\
		\langle k_{\perp}^2(x)\rangle_{g_L}&\in [0.10,0.24] \quad \text{GeV}^2 ,
	\end{align}
	from Table \ref{tb2}, it is evident that although $\langle k_{\perp}^{1,2}(x)\rangle_{g_L}\neq \langle k_{\perp}^{1,2}(x)\rangle_{h_L^{\perp}}$, the equality holds for $\langle k_{\perp}^n\rangle_{g_L}$ and $\langle k_{\perp}^n\rangle_{h_L^{\perp}}$.
	

	\subsection{PDFs}\label{Aqq}
	Integrating the results of $\langle\Gamma\rangle_{\bm{S}}^{(\lambda)}(x,\bm{k}_{\perp})$ over $\bm{k}_{\perp}$ one can obtain the four PDFs of a spin-1 target:
	\begin{subequations}\label{re}
		\begin{align}\label{a93}
			\langle\gamma^+\rangle_{\bm{S}}^{(\lambda)}(x)&\equiv f(x)+S_{LL}f_{LL}(x)\,,\\
			\langle\gamma^+\gamma_5\rangle_{\bm{S}}^{(\lambda)}(x)&\equiv \lambda S_{L}g(x)\,,\\
			\langle\gamma^+\gamma^i\gamma_5\rangle_{\bm{S}}^{(\lambda)}(x)&\equiv \lambda S_Th(x).
		\end{align}
	\end{subequations}
	Integrating Eq. (\ref{tmdj}), we can obtain
	\begin{subequations}\label{tmdf1}
		\begin{align}\label{pc9}
			f(x) &= \frac{N_L^2}{4\pi^2}\int \mathrm{d}\tau \frac{1}{\tau}(3x \bar{x}m_{\rho}^2+ 4 \tau x^2 \bar{x}^2m_{\rho}^4)e^{-\tau B}\nonumber\\
			&+ \frac{N_L^2}{4\pi^2}\int \mathrm{d}\tau \frac{1}{\tau}(M^2+\frac{1}{\tau})e^{-\tau B}\nonumber\\
			&+\frac{N_T^2}{4\pi^2}\int \mathrm{d}\tau  (x^2+\bar{x}^2)(\frac{1}{\tau}+ \bar{x} x m_{\rho}^2)e^{-\tau B}\nonumber\\
			&+\frac{N_T^2}{4\pi^2}\int \mathrm{d}\tau  (1+2x\bar{x} M^2)e^{-\tau B}\,, \\
			g(x)&=\frac{3N_T^2}{8\pi^2}\int \mathrm{d}\tau e^{-\tau B}\nonumber\\
			&\times ((2x-1)(\frac{1}{\tau}-(M^2-\bar{x}x m_{\rho}^2 ))+ M^2)\,, \\
			h(x)&=\frac{3N_TN_LM }{4\sqrt{2}\pi^2}\int \mathrm{d}\tau (\frac{1}{\tau}+2  x \bar{x}m_{\rho}^2)e^{-\tau B}\,, \\
			f_{LL}(x)&=\frac{3N_L^2}{4\pi^2}\int \mathrm{d}\tau  \frac{1}{\tau}(3x \bar{x}m_{\rho}^2+ 4\tau x^2 \bar{x}^2m_{\rho}^4)e^{-\tau B}\nonumber\\
			&+\frac{3N_L^2}{4\pi^2}\int \mathrm{d}\tau  \frac{1}{\tau}(M^2+\frac{1}{\tau})e^{-\tau B}\nonumber\\
			&-\frac{3N_T^2}{8\pi^2}\int \mathrm{d}\tau (x^2+\bar{x}^2)(\frac{1}{\tau}+ \bar{x} x m_{\rho}^2)e^{-\tau B}\nonumber\\
			&-\frac{3N_T^2}{8\pi^2}\int \mathrm{d}\tau (1+2x\bar{x} M^2)e^{-\tau B} \,,
		\end{align}
	\end{subequations}
	where $B=M^2+x(x-1)m_{\rho}^2$.
	\begin{figure}
		\centering
		\includegraphics[width=0.5\textwidth]{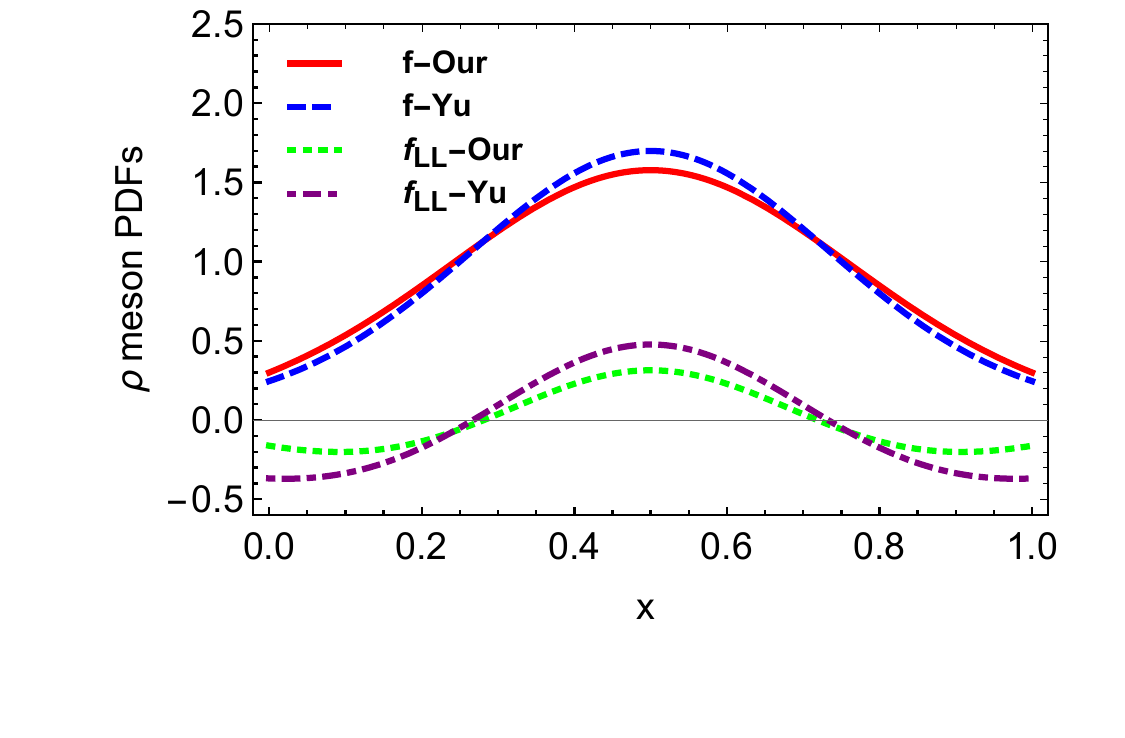}
		\qquad
		\includegraphics[width=0.51\textwidth]{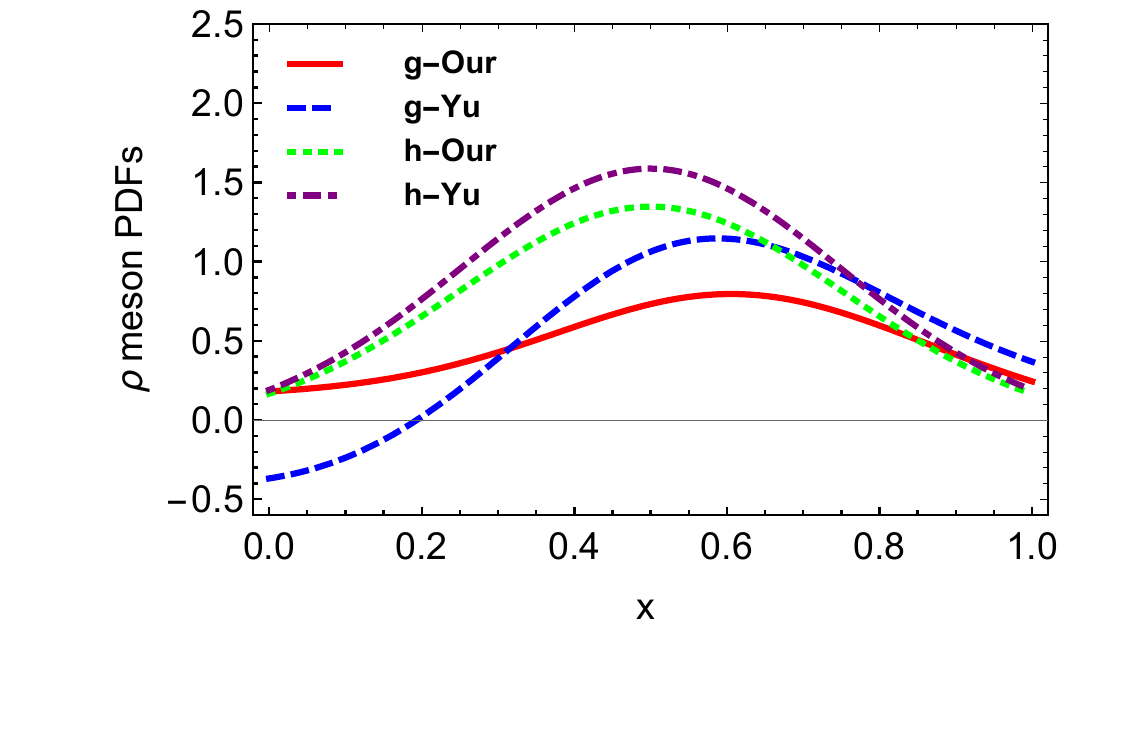}
		\caption{The unpolarized, tensor-polarized, helicity, and transversity PDFs of the $\rho$ meson presented in this paper at the model scale $Q_0^2=0.16$ GeV$^2$ are compared with the results obtained by Yu Ninomiya in Ref.~\cite{Ninomiya:2017ggn} at the same model scale. }\label{pdf1}
	\end{figure}
	Note that the tensor polarized PDF $f_{LL}(x)$ is related to the single flavor distribution function $b_1(x)$,
	\begin{align}\label{a93}
		b_1(x)=\frac{1}{2}f_{LL}(x),
	\end{align}
	$b_1(x)$ measures the difference in the spin projection of the $\rho$ meson, which depends solely on the quark-spin-averaged distribution
	
	The PDFs of $\rho$ meson are depicted in Fig. \ref{pdf1}. A comparison with Ref.~\cite{Ninomiya:2017ggn} reveals that $f(x)$ and $f_{LL}(x)$ exhibit similarity, while the numerical value of our $h(x)$ is slightly smaller than their reported results. The primary disparity lies in the behavior of $g(x)$: in Ref.~\cite{Ninomiya:2017ggn}, $g(x)$ is negative for small values of $x$, whereas our findings indicate positivity across the entire range of $x\in [0,1]$. 
	
	The PDFs contain the information of the longitudinal momentum and the polarization carried by the partons. Here we define the $x$-moments of various PDFs as
	\begin{align}\label{pcc1}
		\langle x^n \rangle_{\alpha}=\int_0^1 \mathrm{d}x x^n \alpha(x),
	\end{align}
	where $\alpha(x)$ is the $\rho$ meson PDF.
	
	The following sum rules should be satisfied,
	\begin{align}\label{pc1}
		\int_0^1 f(x)\mathrm{d}x =1, \quad \int_0^1 f_{LL}(x)\mathrm{d}x=0,
	\end{align}
	which means that in our model we have one valence $u$ quark and one valence $d$ quark in the $\rho$ meson. The sum rule of $f_{LL}$ corresponds to the $b_1$ sum rule in our model. The tensor polarized PDF $f_{LL}$ is symmetric around $x = 1/2$, it is negative in the small and large $x$ region, and positive around the peak at $x = 1/2$, $f_{LL}$ has two nodes at the model scale. $f(x)$ is also symmetric around $x = 1/2$, we obtain the following results for the momentum sum rule of the valence quark:
	\begin{align}\label{pcx}
		\int_0^1 xf(x)\mathrm{d}x =\frac{1}{2}, \quad \int_0^1 xf_{LL}(x)\mathrm{d}x=0,
	\end{align}
	this implies the valence quark and antiquark carry $100\%$ of the light-cone momentum, regardless of the spin quantization axis $S$ or the spin projection $\lambda$ of the hadron. 
	
	The spin sum rule of $g(x)$ is
	\begin{align}\label{pc1}
		\Delta q\equiv\int_0^1 g(x)\mathrm{d}x=0.505,
	\end{align}
	which means the total contribution of the valence quark and antiquark to the spin of the $\rho$ meson is $50.5\%$. In contrast, Ref.~\cite{Ninomiya:2017ggn} reports this value as $ 56\% $, which is higher. 
	
	The sum rules of $h(x)$ is
	\begin{align}\label{pc1}
		\int_0^1 h(x)\mathrm{d}x=0.803,
	\end{align}
	which indicates that $u$ quark of the $\rho$ meson’s tensor charge at the model scale is $0.803$, which is smaller than $0.94$  reported in Ref.~\cite{Ninomiya:2017ggn}. The values are considerably larger than the spin sum and approximate the naive quark model expectation of unity. In Table \ref{tb3}, we list the $\langle x^0\rangle$ and $\langle x^1\rangle$ moments of the various PDFs, and compared with the results in Refs.~\cite{Ninomiya:2017ggn,Shi:2022erw}. Table ~\ref{tb4} compares the $\langle x^n\rangle_g$ moments of the helicity PDF. It shows that for $\langle x^0\rangle_g$, the NJL result from the covariant method is close to the Lattice results. For $\langle x\rangle_g$, our result is a slightly larger than both the BSE and Lattice results. For $\langle x^2\rangle_g$, our result is larger than both the BSE and Lattice result, particularly the Lattice result.

	\begin{center}
		\begin{table}
			\caption{The $\langle x^0\rangle$ and $\langle x^1\rangle$ moments of $g(x)$ and $h(x)$ defined in Eq. (\ref{pcc1}). Here we compare with the results in the NJL model~\cite{Ninomiya:2017ggn}, Ref.~\cite{Shi:2022erw} employs the BSE-based LF-LFWFs and Ref.~\cite{Kaur:2024iwn} utilizes LF quantization.}\label{tb3}
			\begin{tabular}{p{3.5cm}p{1.5cm} p{1.5cm} }
				\hline\hline
				&$\langle x^0\rangle_g$&$\langle x^1\rangle_h$ \\
				\hline
				NJL ~\cite{Ninomiya:2017ggn}&0.56&0.94\\
				\hline
				BSE ~\cite{Shi:2022erw}&0.67&0.79\\
				\hline
				LF quantization~\cite{Kaur:2024iwn}&0.54&0.70\\
				\hline
				This work &0.505&0.803\\
				\hline\hline
			\end{tabular}
		\end{table}
	\end{center}

	\begin{center}
		\begin{table}
			\caption{Comparison of $\langle x^n\rangle_g$ moments of the helicity PDF.}\label{tb4}
			\begin{tabular}{p{3.0cm}p{1.7cm} p{1.7cm}p{1.7cm} }
				\hline\hline
				&$\langle x^0\rangle_g$&$\langle x\rangle_g$&$\langle x^2\rangle_g$ \\
				\hline
				NJL ~\cite{Ninomiya:2017ggn}&0.558&0.369&0.251\\
				\hline
				BSE ~\cite{Shi:2022erw}&0.670&0.227&0.111\\
				\hline
				Lattice QCD~\cite{Best:1997qp}&0.570(32)&0.212(17)&0.077(34)\\
				\hline
				This work &0.505&0.279&0.183\\
				\hline\hline
			\end{tabular}
		\end{table}
	\end{center}

	\subsection{Positivity constraints}\label{Aqq}
\begin{figure*}
		\centering
		\includegraphics[width=0.32\textwidth]{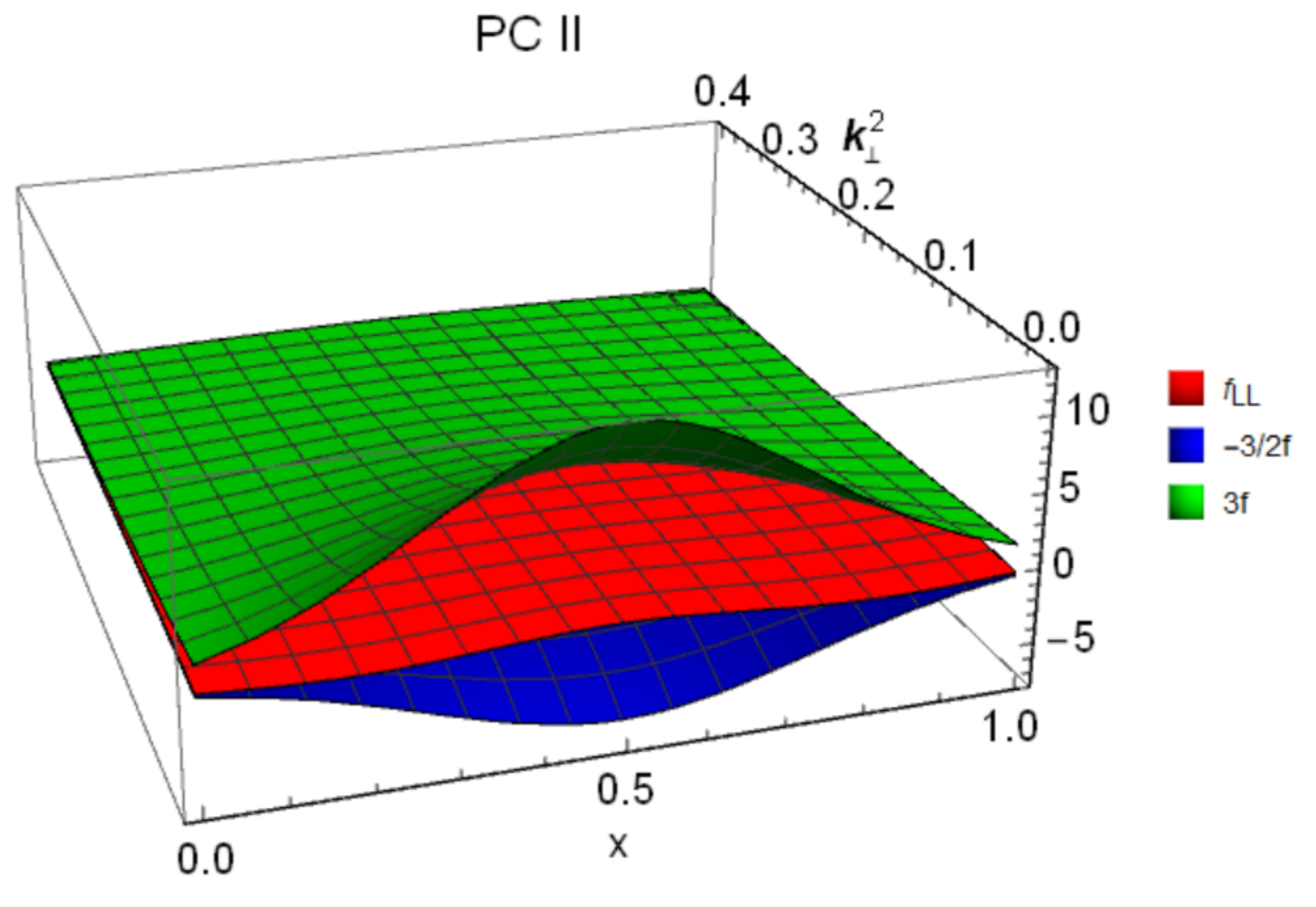}
		\includegraphics[width=0.32\textwidth]{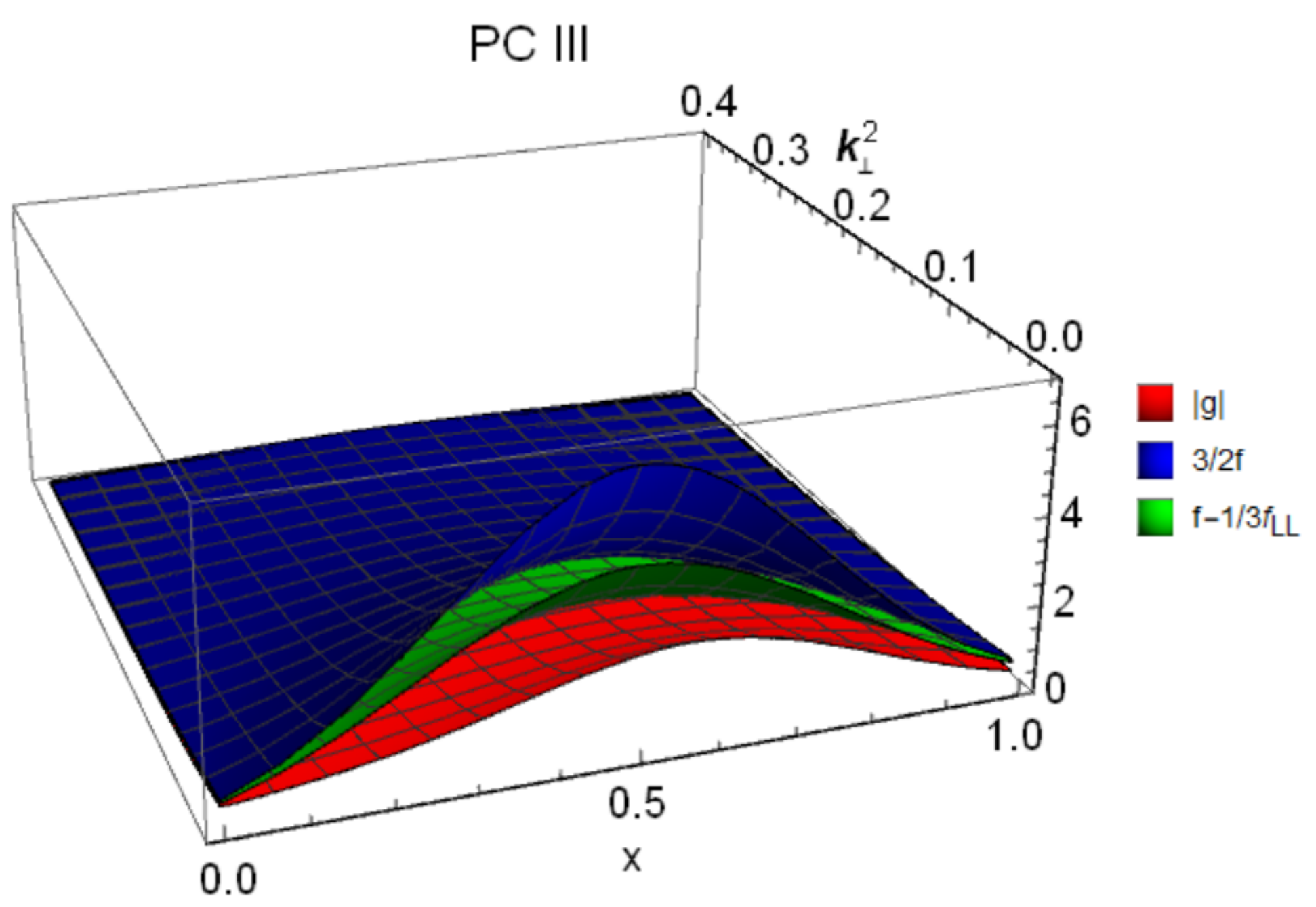}
		\includegraphics[width=0.32\textwidth]{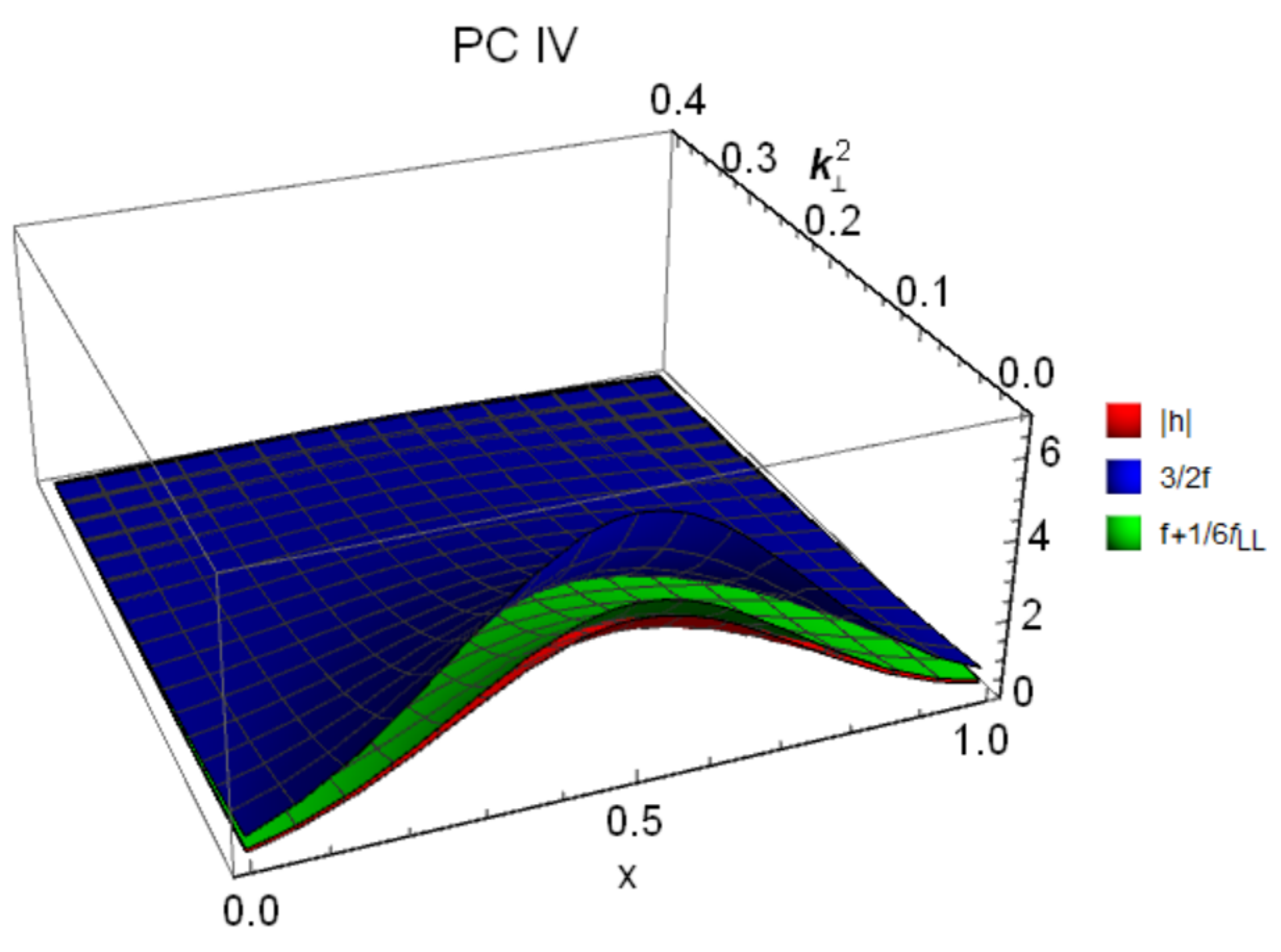}
		\caption{The positivity constraints for the quark TMDs of the $\rho$ meson expressed in Eq. (\ref{tmdpc6}). }\label{tmdpc}
	\end{figure*}

	From the probabilistic interpretation $\langle\Gamma \rangle_{\bm{S}}^{(\lambda)}(x,\bm{k}_{\perp}^2)$, given by the quantities, one can obtain the inequalities:
	\begin{align}\label{pc4}
		\langle \gamma^+ \rangle_{\bm{S}}^{(\lambda)}(x,\bm{k}_{\perp}^2)\geqslant 0,
	\end{align}
	\begin{align}\label{pc6}
		\langle \gamma^+ \rangle_{\bm{S}}^{(\lambda)}(x,\bm{k}_{\perp}^2)\geqslant |\langle \gamma^+\gamma^5 \rangle_{\bm{S}}^{(\lambda)}(x,\bm{k}_{\perp}^2)|,
	\end{align}
	\begin{align}\label{pc7}
		\langle \gamma^+ \rangle_{\bm{S}}^{(\lambda)}(x,\bm{k}_{\perp}^2)\geqslant |\langle \gamma^+\gamma^i\gamma^5 \rangle_{\bm{S}}^{(\lambda)}(x,\bm{k}_{\perp}^2)|.
	\end{align}
	These relations must be valid for any spin quantization axis $\bm{S}$ and spin projection $\lambda$. Thus the spin-$1$ hadron, TMDs should satisfy the following relations ~\cite{Ninomiya:2017ggn,Bacchetta:2001rb}

	\begin{subequations}\label{tmdpc6}
		\begin{align}
			&\text{I}. \quad f(x,\bm{k}_{\perp}^2) \geqslant 0\,, \\
			&\text{II}.\quad -\frac{3}{2}f(x,\bm{k}_{\perp}^2)\leqslant f_{LL}(x,\bm{k}_{\perp}^2)\leqslant 3f(x,\bm{k}_{\perp}^2)\,, \\
			&\text{III}.\quad |g_L (x,\bm{k}_{\perp}^2)| \leqslant f(x,\bm{k}_{\perp}^2) -\frac{1}{3} f_{LL}(x,\bm{k}_{\perp}^2)\leqslant \frac{3}{2}f(x,\bm{k}_{\perp}^2)\,, \\
			&\text{IV}.\quad |h(x,\bm{k}_{\perp}^2)|\leqslant f(x,\bm{k}_{\perp}^2) +\frac{1}{6} f_{LL}(x,\bm{k}_{\perp}^2)\leqslant \frac{3}{2}f(x,\bm{k}_{\perp}^2).
		\end{align}
	\end{subequations}
	For the first constraint, the first diagram in Fig. \ref{tmd1} shows that our results satisfy the condition $f(x,\bm{k}_{\perp}) \geqslant 0$. The remaining positivity constraints are illustrated in Fig. \ref{tmdpc}. These diagrams demonstrate that our $\rho$ meson TMDs fulfill the positivity constraints I–IV.

\begin{figure*}
		\centering
		\includegraphics[width=0.32\textwidth]{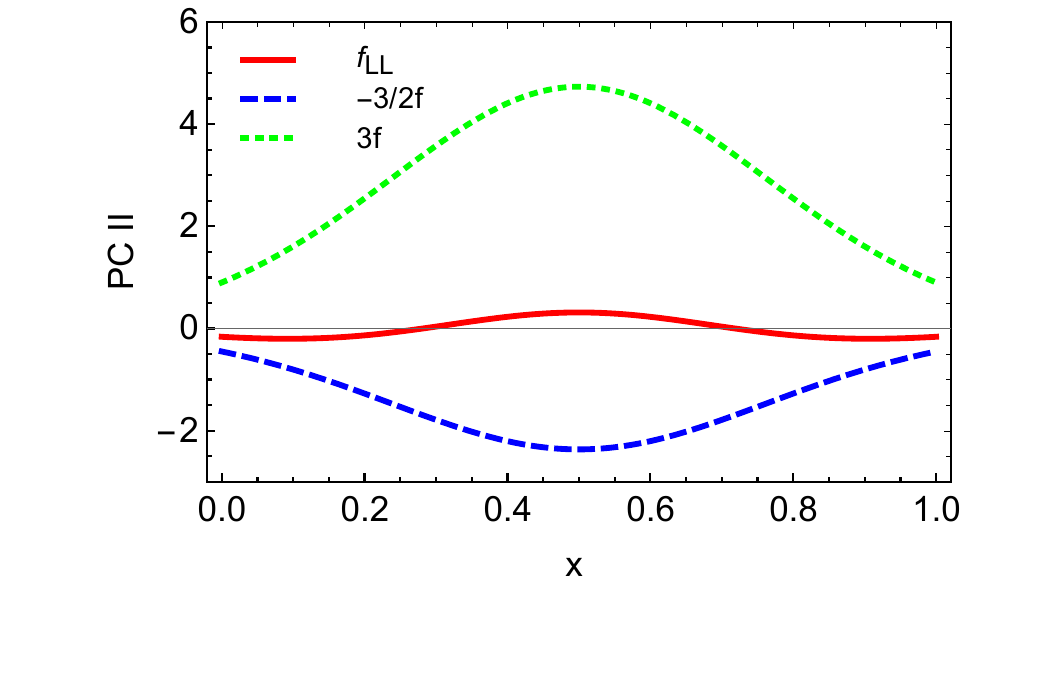}
		\includegraphics[width=0.32\textwidth]{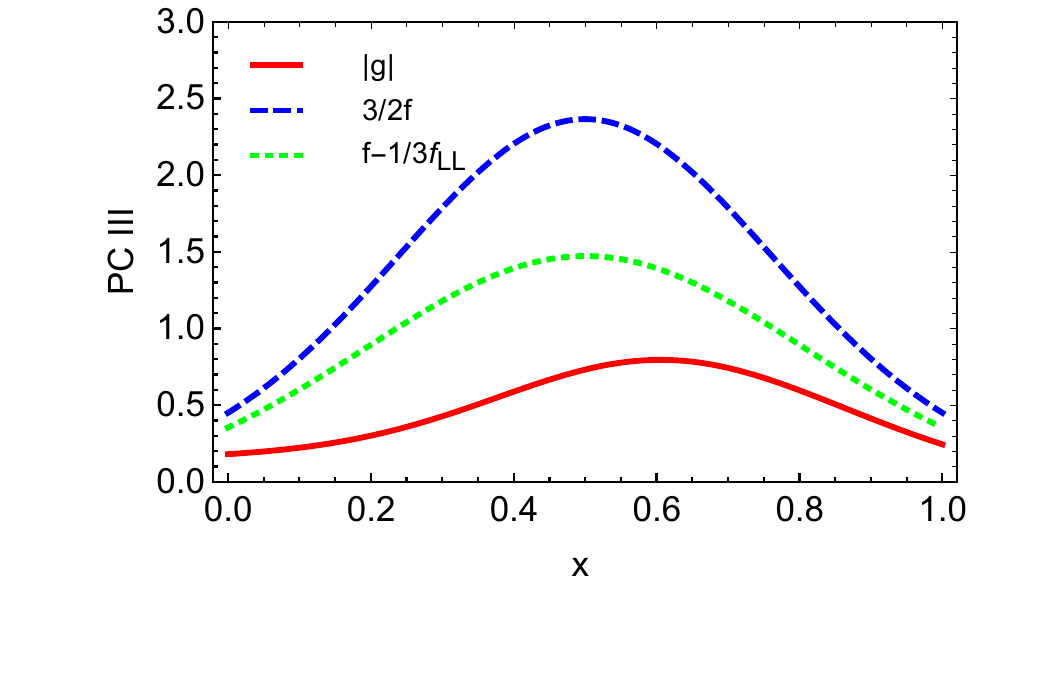}
		\includegraphics[width=0.32\textwidth]{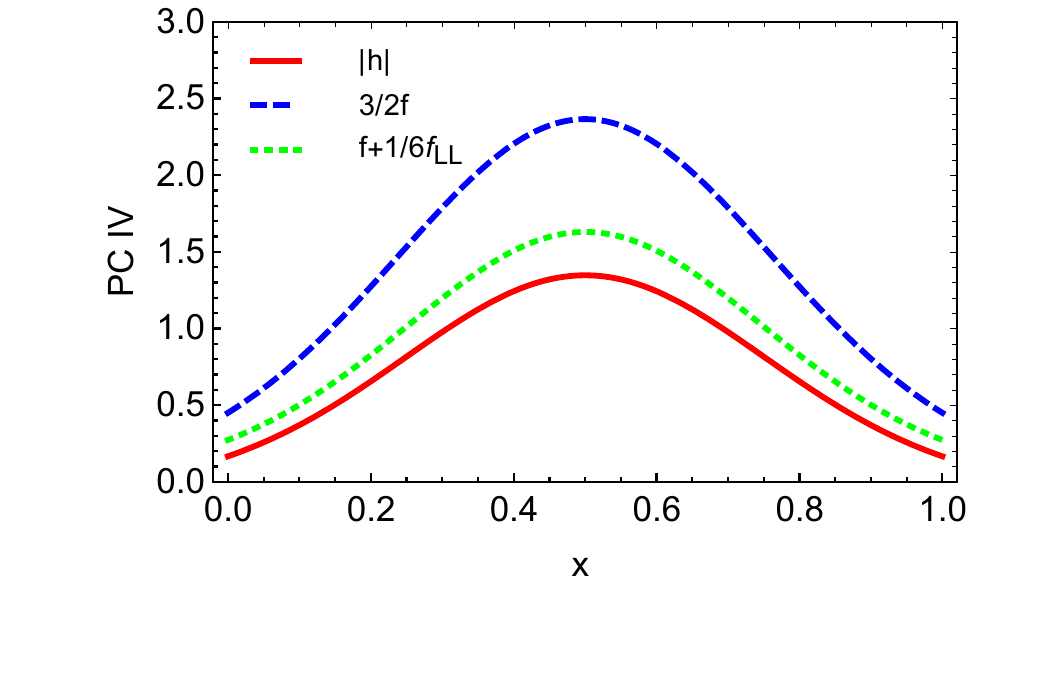}
		\caption{The positivity constraints for the quark PDFs of the $\rho$ meson expressed in Eq. (\ref{pc10}). }\label{pc}
	\end{figure*}

	Integrating TMDs equations above over $\bm{k}_{\perp}$ gives the following naive positivity conditions on the PDFs:
	\begin{subequations}\label{pc10}
		\begin{align}
			&\text{I}. \quad \quad  f(x) \geqslant 0\,, \\
			&\text{II}. \quad \quad  -\frac{3}{2}f(x)\leqslant f_{LL}(x)\leqslant 3f(x)\,, \\
			&\text{III}.  \quad \quad  |g_L (x)| \leqslant f(x) -\frac{1}{3} f_{LL}(x)\leqslant \frac{3}{2}f(x)\,, \\
			&\text{IV}. \quad \quad  |h(x)|\leqslant f(x)+\frac{1}{6} f_{LL}(x)\leqslant \frac{3}{2}f(x).
		\end{align}
	\end{subequations}
	In Fig. \ref{pc}, we show that our results satisfy the positivity constraints effectively. From Fig. \ref{pdf1}, we observe that $f(x)>0$, indicating that the first condition is also met. Both Figs. \ref{pdf1} and \ref{pc} demonstrate that our $\rho$ meson PDFs satisfy constraints I–IV. In contrast to the findings in Ref.~\cite{Ninomiya:2017ggn}, our results reveal deviations from the relationships $f_{LL}=-\frac{3}{2}f$, $|g|=\frac{3}{2}f$ and $|h|=\frac{M}{m_{\rho}}\frac{3}{2}f$ at the endpoints $x=0$ and $x=1$, as shown in Fig. \ref{pc}.
	%

	\subsection{Spin densities in the momentum space}
	The TMDs can be explained as the quark densities inside the hadron. The distribution of quark momentum inside the target can be defined by various polarization combinations represented through TMDs. Follow Ref.~\cite{Kaur:2020emh}, the quark spin densities in the momentum space for the spin-1 target are defined as,
	
	\begin{widetext}
		\begin{align}\label{sd}
			&\rho(x,k_x,k_y,(\lambda,\bm{\lambda}_{\perp}),(\Lambda,\bm{\Lambda}_{\perp}))=f+\lambda \Lambda g_L +\lambda \Lambda_{\perp}^i\frac{k_{\perp}^i}{m_{\rho}}g_T+\lambda_{\perp}^i\Lambda_{\perp}^ih+\lambda_{\perp}^i \Lambda \frac{k_{\perp}^i}{m_{\rho}} h_L^{\perp}\nonumber\\
			&+(3\lambda^2-2)\left(\left(\frac{1}{6}-\frac{1}{2}\Lambda^2\right)f_{LL}+\Lambda\Lambda_{\perp}^i\frac{k_{\perp}^i}{m_{\rho}}f_{LT}+\left(\Lambda_{\perp}^i\Lambda_{\perp}^j-\frac{1}{2}\Lambda_{\perp}^2\delta_{ij}\right)\frac{k_{\perp}^ik_{\perp}^j}{m_{\rho}^2}f_{TT} \right),
		\end{align}
	\end{widetext}
	in the longitudinal direction, $\lambda$ and $\Lambda$ denote the spins of the quark and the target, respectively. These symbols represent different quantities from previous sections: $\lambda=\uparrow,\downarrow $ (or $+1$, $-1$) and $\Lambda=\uparrow,\downarrow $ (or $+1$, $-1$). The transverse spins of the quark and the target $\rho$ meson are denoted as $\bm{\lambda}_{\perp}=\Uparrow,\Downarrow$ (or $+1$, $-1$), and $\bm{\Lambda}_{\perp}=\Uparrow,\Downarrow$ (or $+1$, $-1$), respectively. In this paper, we consider transverse polarization along the $x$-direction.
	
\begin{figure*}
		\centering
		\includegraphics[width=0.3\textwidth]{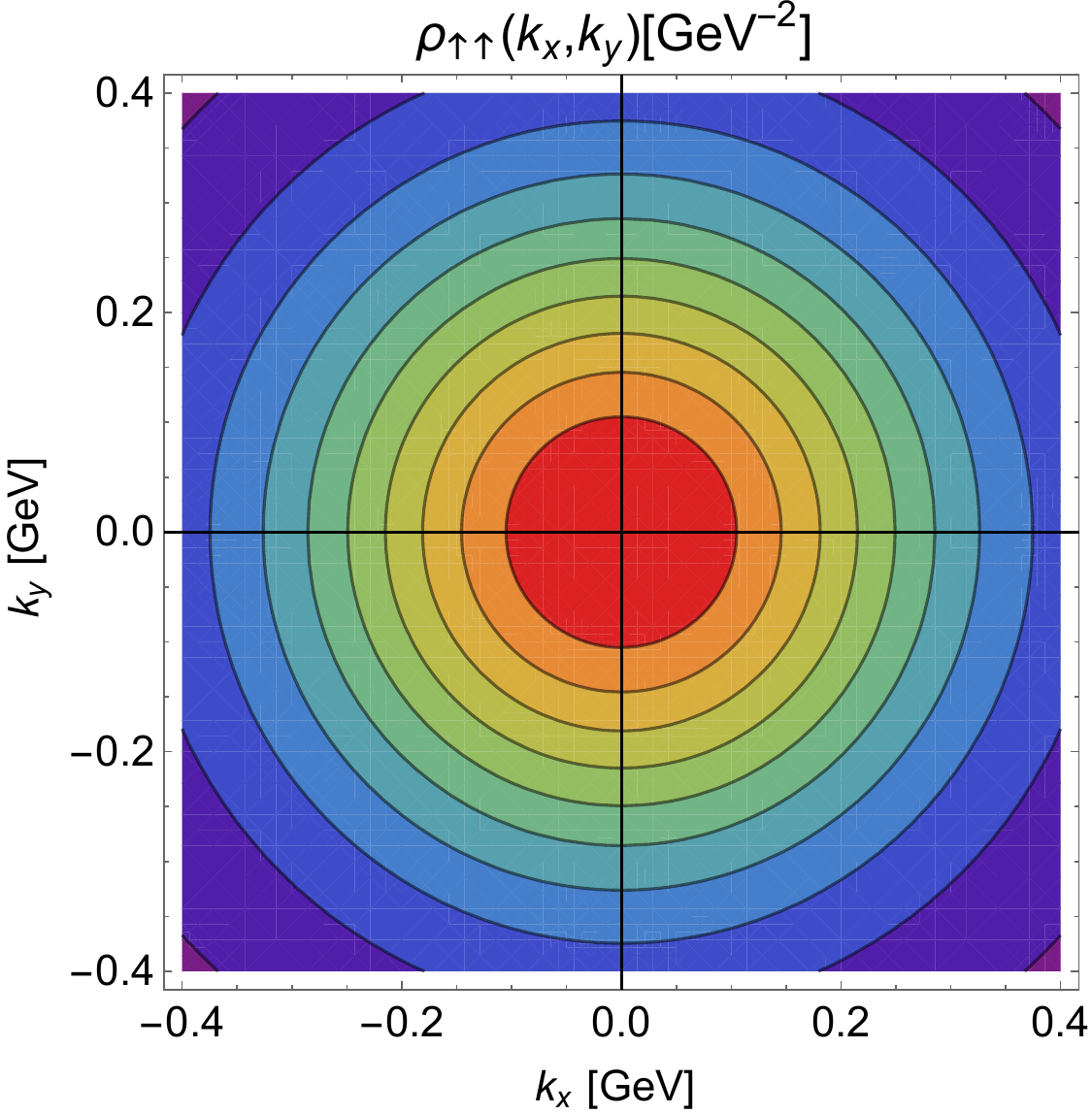}
		\includegraphics[width=0.038\textwidth]{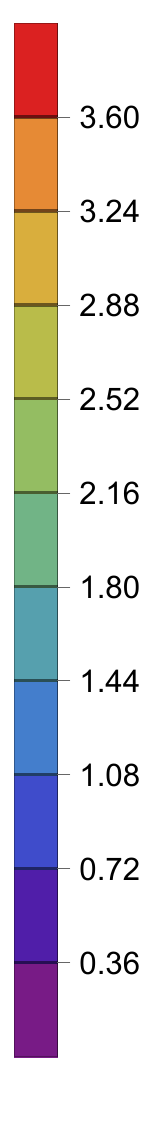}
		\qquad\qquad\qquad
		\includegraphics[width=0.3\textwidth]{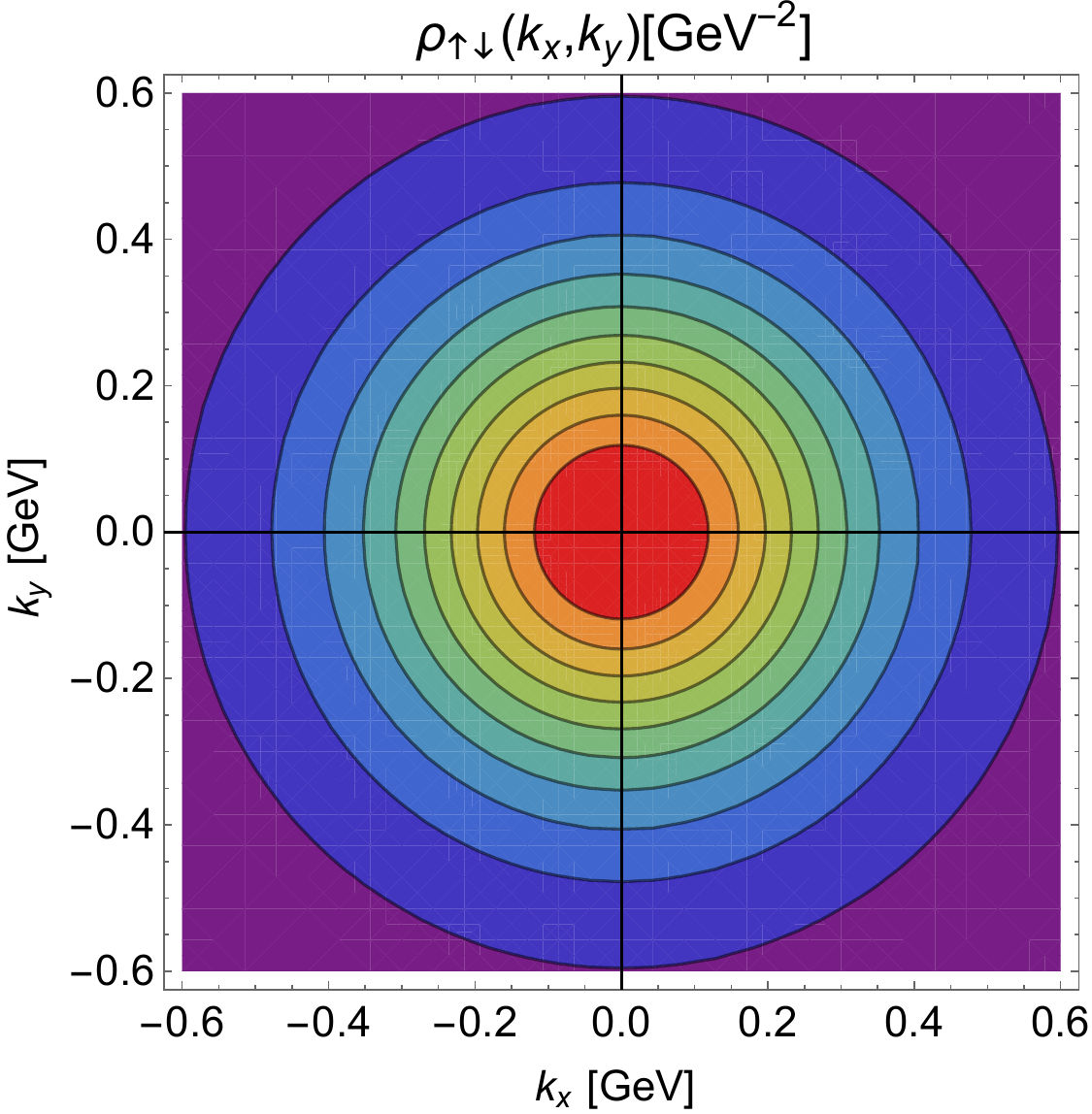}
		\includegraphics[width=0.033\textwidth]{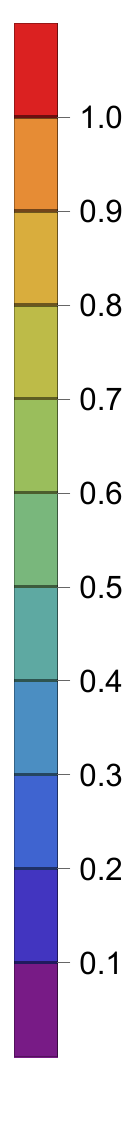}
		\qquad\qquad\qquad
		\includegraphics[width=0.3\textwidth]{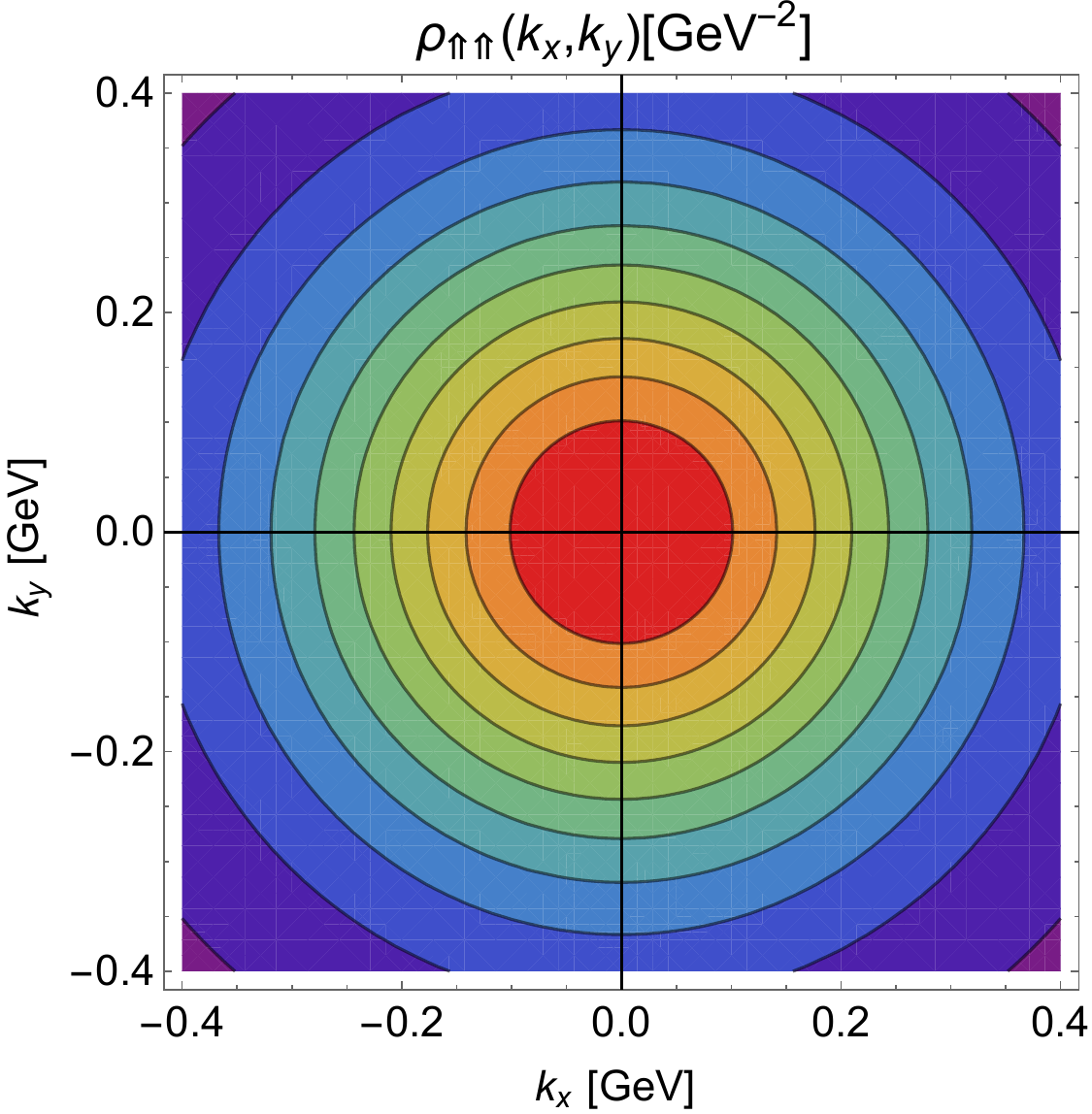}
		\includegraphics[width=0.038\textwidth]{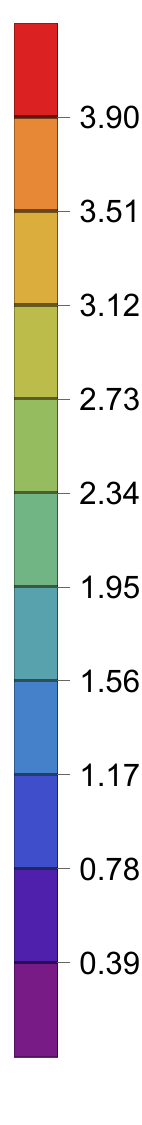}
		\qquad\qquad\qquad
		\includegraphics[width=0.3\textwidth]{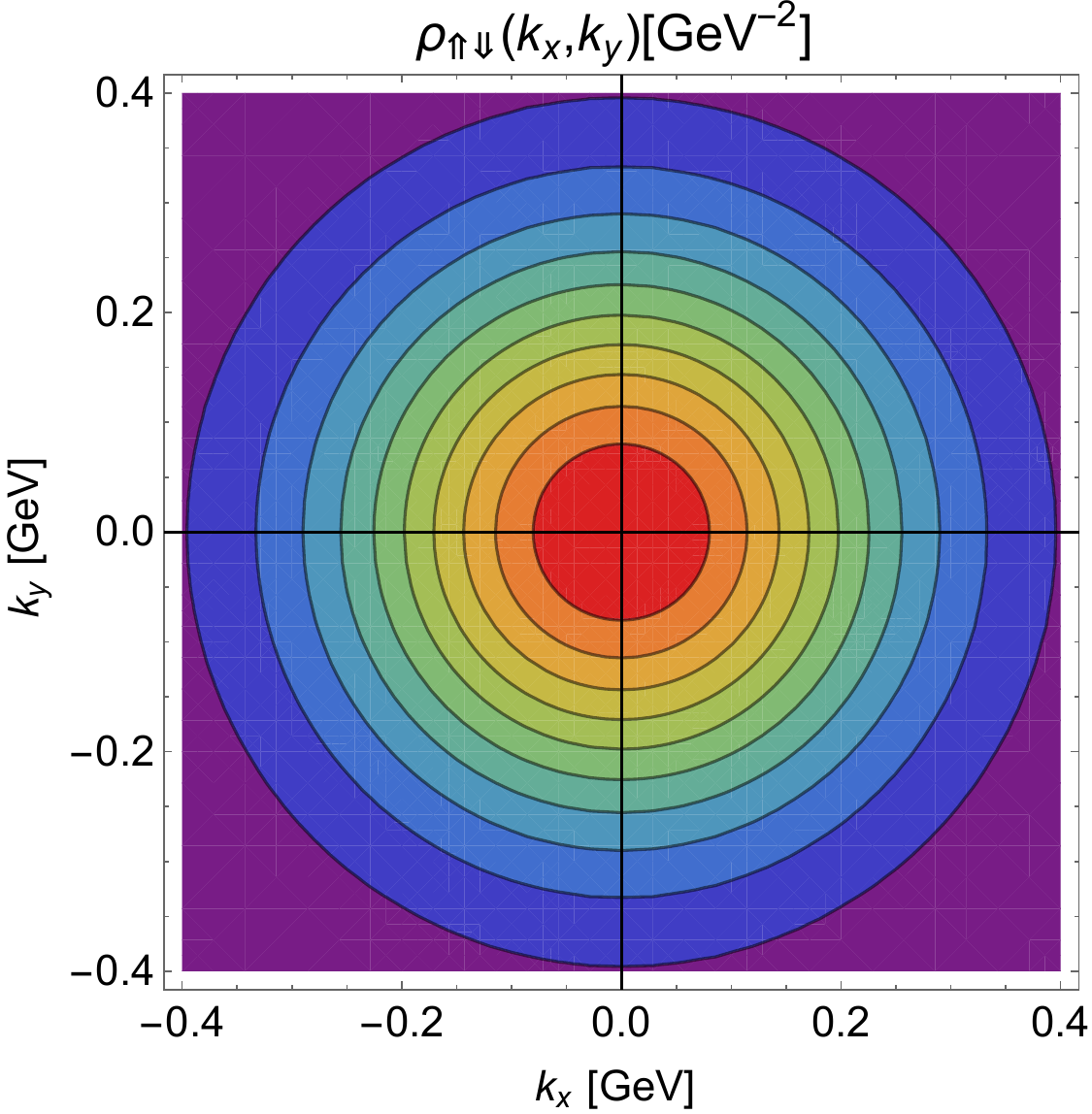}
		\includegraphics[width=0.043\textwidth]{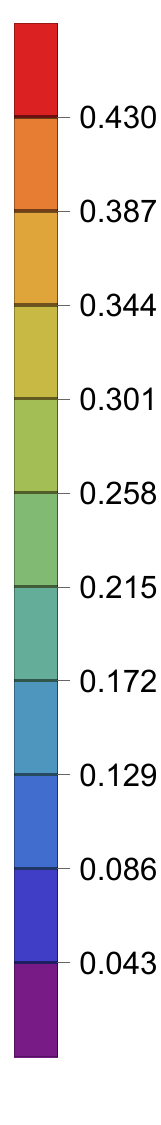}
		\qquad\qquad\qquad
		\includegraphics[width=0.3\textwidth]{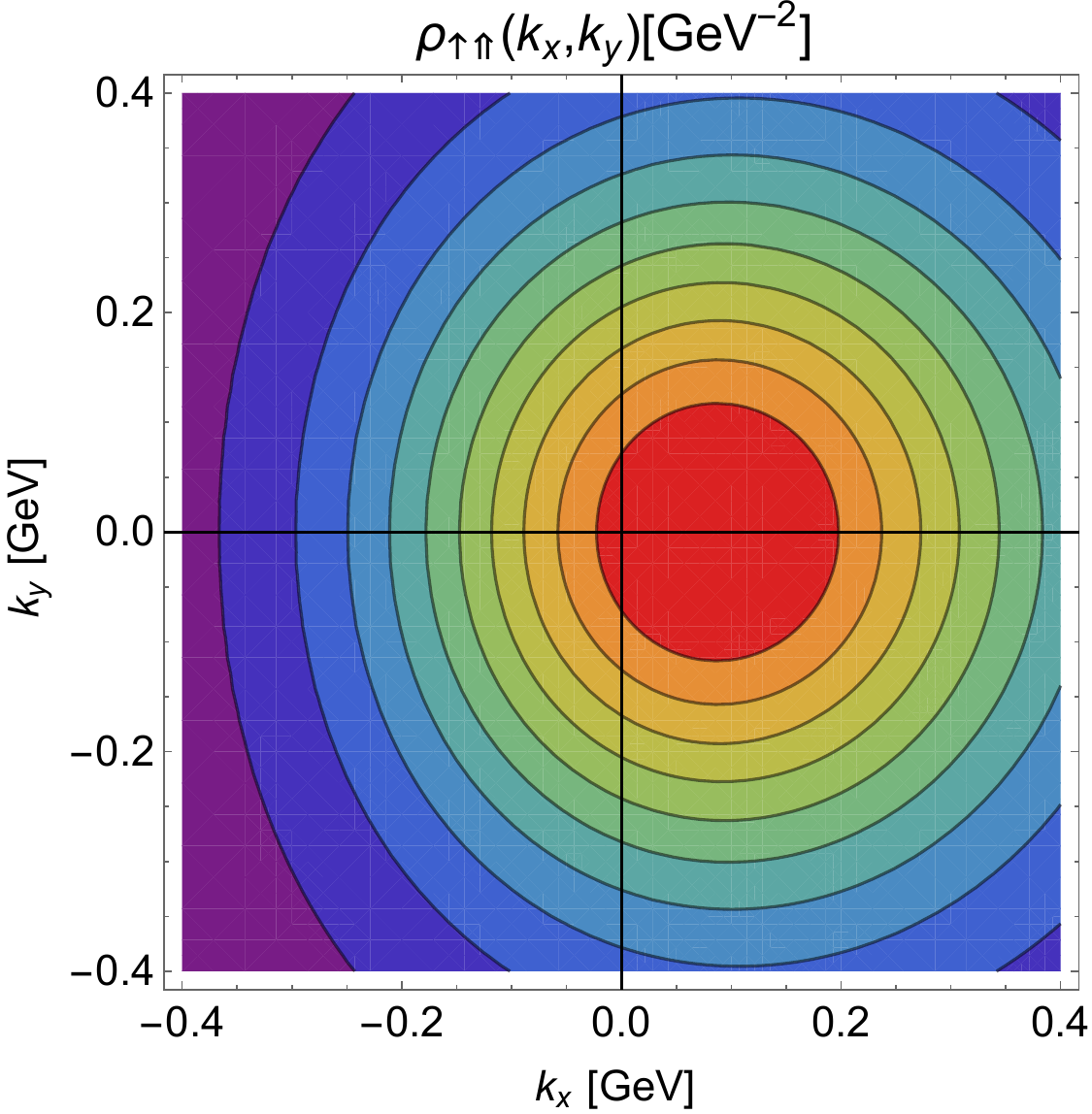}
		\includegraphics[width=0.038\textwidth]{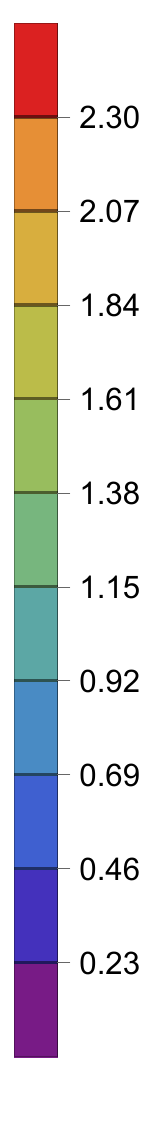}
		\qquad\qquad\qquad
		\includegraphics[width=0.3\textwidth]{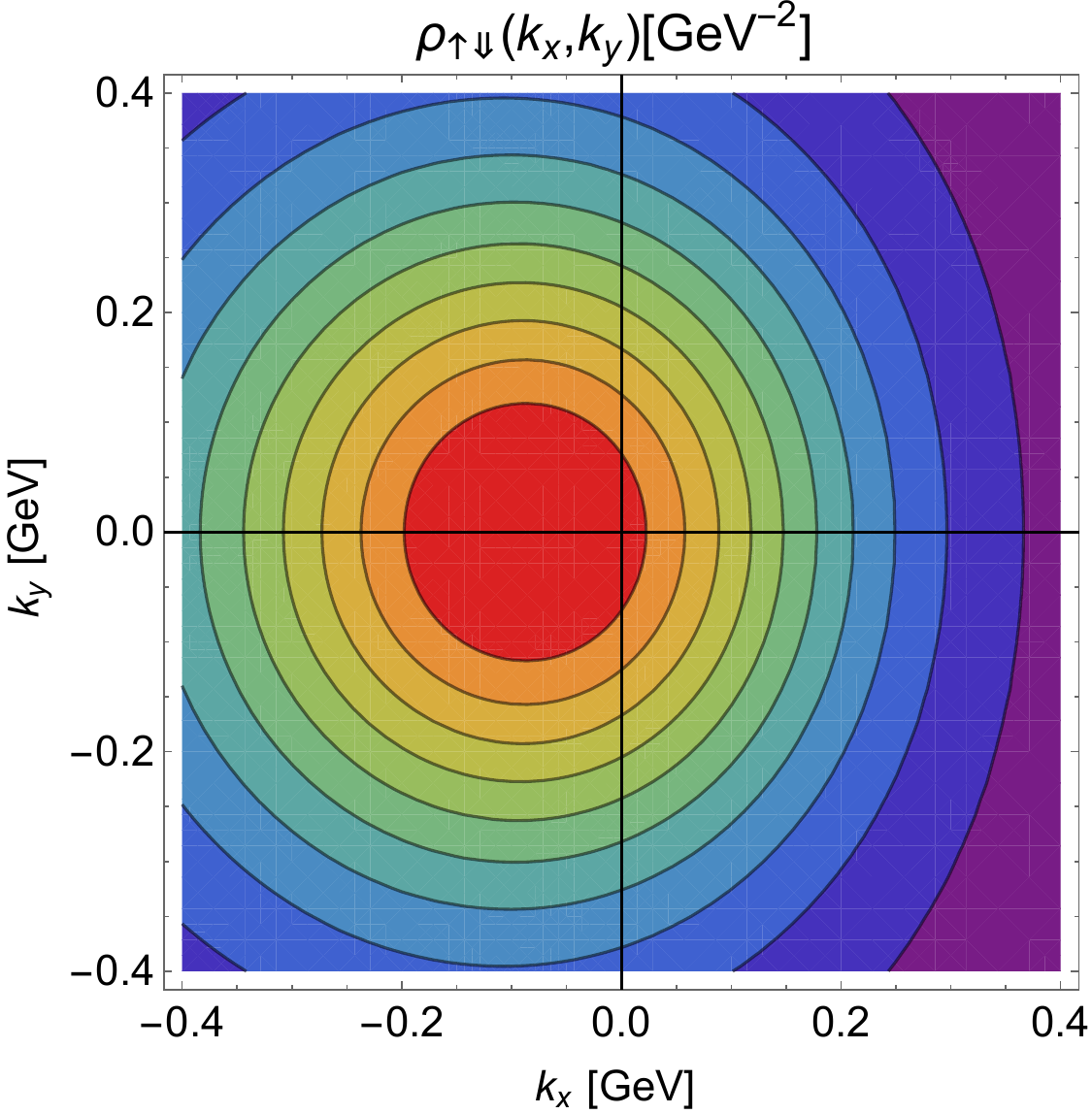}
		\includegraphics[width=0.038\textwidth]{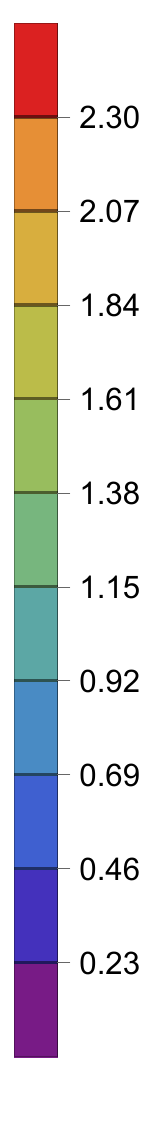}
		\qquad\qquad\qquad
		\includegraphics[width=0.3\textwidth]{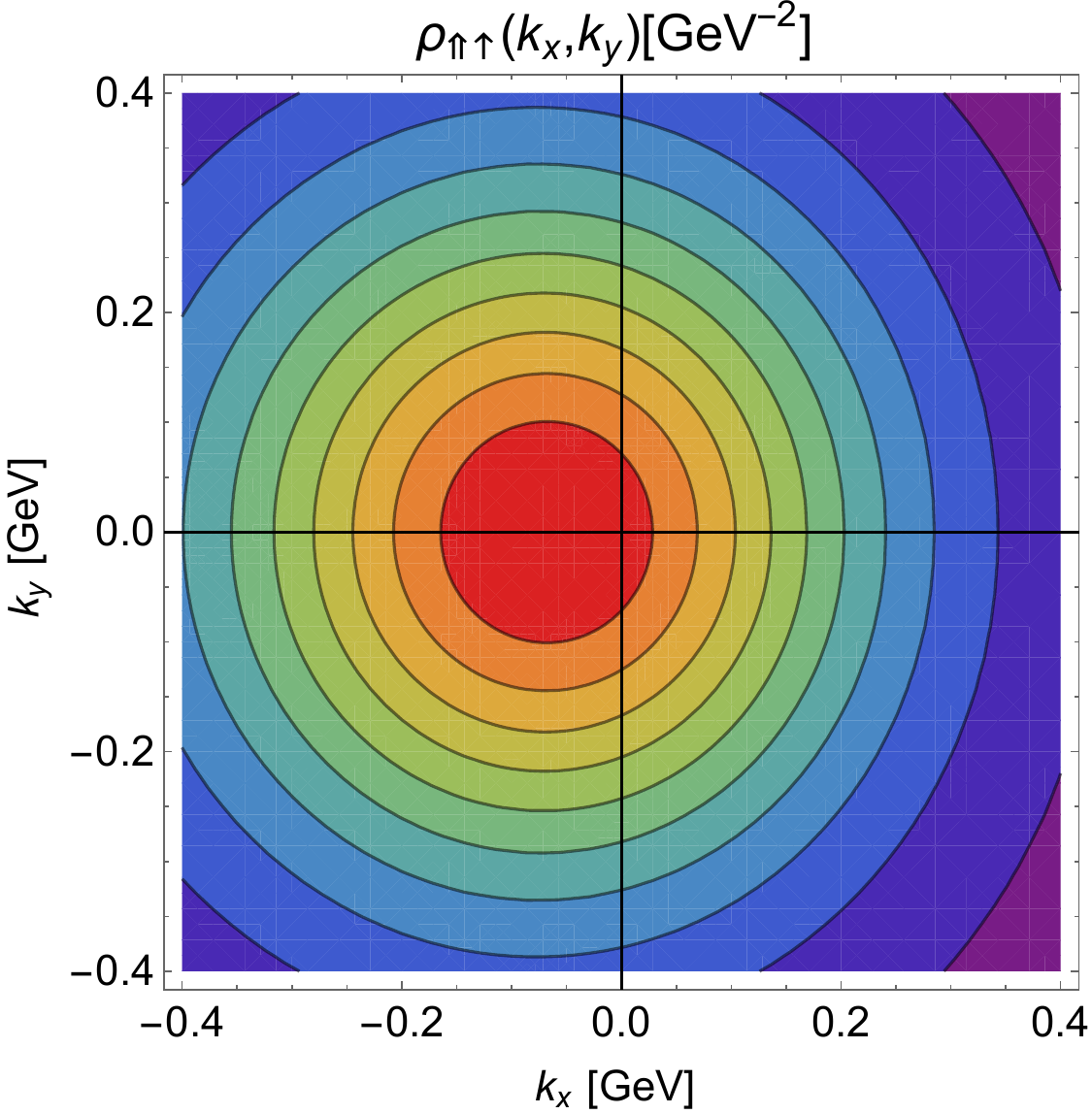}
		\includegraphics[width=0.038\textwidth]{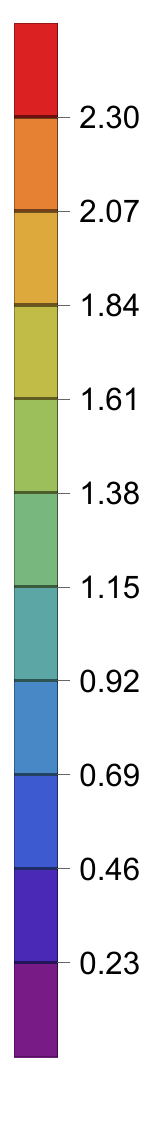}
		\qquad\qquad\qquad
		\includegraphics[width=0.3\textwidth]{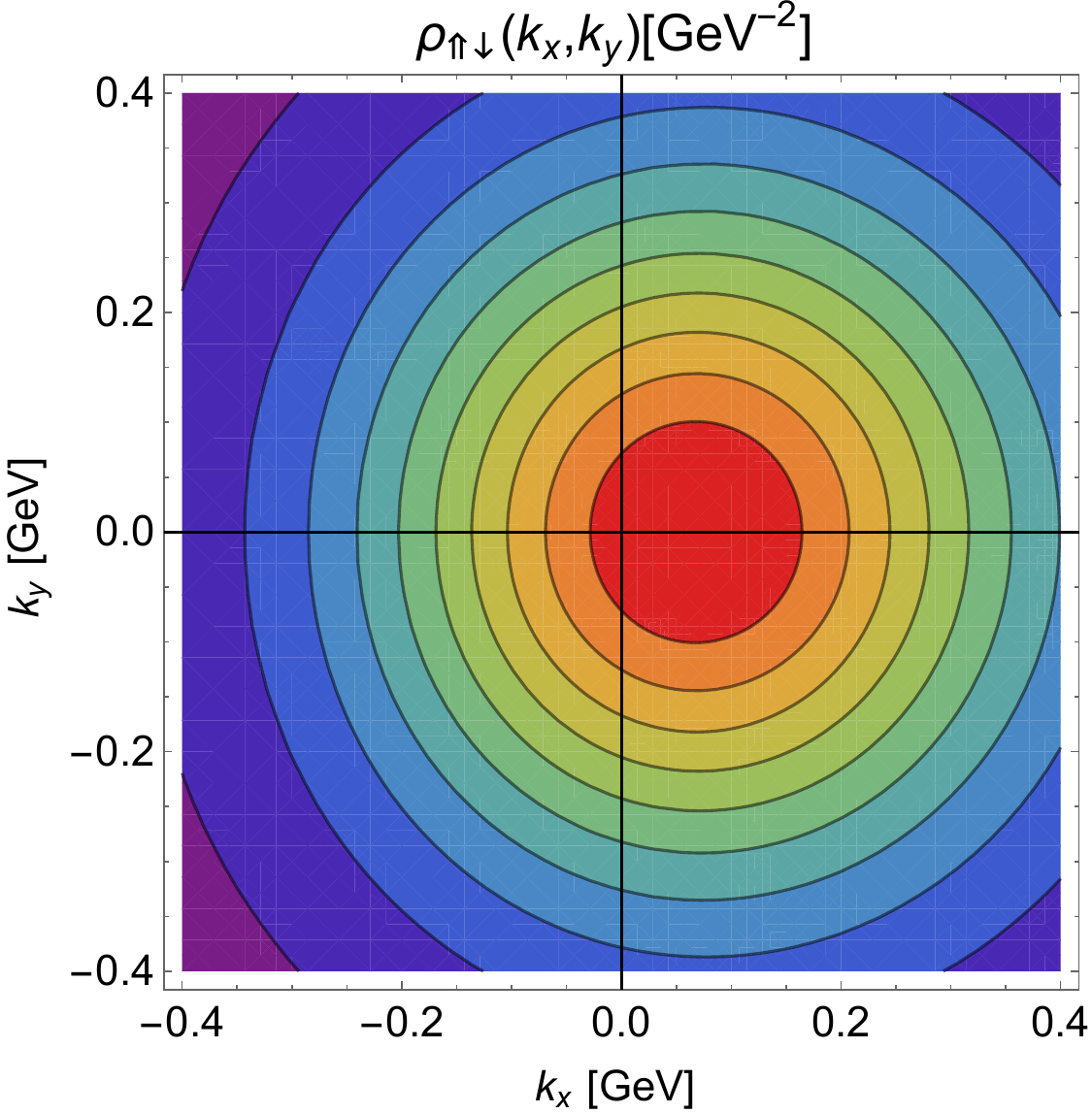}
		\includegraphics[width=0.038\textwidth]{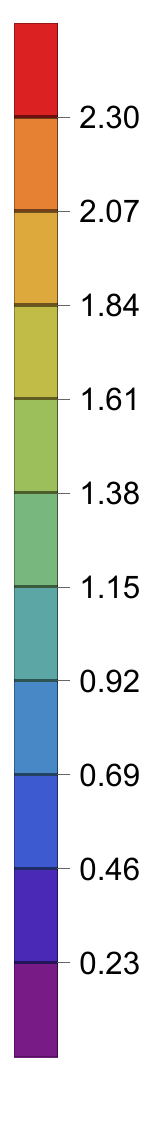}
		\caption{The $u$ quark spin densities of the $\rho$ meson in the momentum space: $\rho_{\uparrow\uparrow}(k_x,k_y)$ and $\rho_{\uparrow\downarrow}(k_x,k_y)$ in the first row; $\rho_{\Uparrow\Uparrow}(k_x,k_y)$ and $\rho_{\Uparrow\Downarrow}(k_x,k_y)$ in the second row; $\rho_{\uparrow\Uparrow}(k_x,k_y)$ and $\rho_{\uparrow \Downarrow}(k_x,k_y)$ in the third row; $\rho_{\Uparrow\uparrow}(k_x,k_y)$ and $\rho_{\Uparrow\downarrow}(k_x,k_y)$ in the fourth row;}\label{kaondp}
	\end{figure*}

	To obtain all spin densities in the transverse momentum plane, we performed an integration of the longitudinal momentum fraction $x$. From Eq. (\ref{sd}) one can obtain $\rho_{\uparrow\uparrow}$ and $\rho_{\uparrow\downarrow}$
	\begin{align}\label{sd1}
		\rho_{\uparrow\uparrow}(k_x,k_y)&=f+g_L-\frac{1}{3}f_{LL}\,,\\
		\rho_{\uparrow\downarrow}(k_x,k_y)&=f-g_L-\frac{1}{3}f_{LL}\,,
	\end{align}
	$\rho_{\downarrow\downarrow}=\rho_{\uparrow\uparrow}$ and $\rho_{\downarrow\uparrow}=\rho_{\uparrow\downarrow}$, therefore, our focus will be solely on studying $\rho_{\uparrow\uparrow}$ and $\rho_{\uparrow\downarrow}$. $\rho_{\uparrow\uparrow}$ represents the likelihood of locating the quark within the $\rho$ meson with its spin aligned to the spin of the composite system. $\rho_{\uparrow\uparrow}$ denotes the probability of both spins being in an anti-aligned state. The spin densities of the quarks in the momentum space for the $\rho$ meson are depicted in Fig. \ref{kaondp}. In the first row, we have $\rho_{\uparrow\uparrow}$ and $\rho_{\uparrow\downarrow}$, both exhibiting an axially symmetric behavior in transverse momentum space. It is worth noting that $\rho_{\uparrow\uparrow}$ is larger than $\rho_{\uparrow\downarrow}$ due to the opposite signs of $f$ and $g_L$. Additionally, it is observed that while the covariant approach in Ref.~\cite{Ninomiya:2017ggn} yields a negative value for $\rho_{\uparrow\downarrow}$, our calculation results in a positive value for this quantity.

	$\rho_{\Uparrow\Uparrow}$ and $\rho_{\Uparrow,\Downarrow}$ are defined as
	\begin{align}\label{sd2}
		\rho_{\Uparrow\Uparrow}(k_x,k_y)&=f+h+\frac{\bm{k}_{\perp}^2}{m_{\rho}^2}f_{TT}\,,\\
		\rho_{\Uparrow\Downarrow}(k_x,k_y)&=f-h+\frac{\bm{k}_{\perp}^2}{m_{\rho}^2}f_{TT},
	\end{align}
	and $\rho_{\Downarrow\Downarrow}=\rho_{\Uparrow\Uparrow}$ and $\rho_{\Downarrow\Uparrow}=\rho_{\Uparrow\Downarrow}$. $\rho_{\Uparrow\Uparrow}$ and $\rho_{\Uparrow\Downarrow}$ denote the combination of zero and two units of orbital angular momentum (OAM) transfer overlap terms. They are depicted in the second row of Fig. \ref{kaondp}. It can be observed that they also exhibit axial symmetry in the transverse momentum space, with the numerical value of $\rho_{\Uparrow\Uparrow}$ being greater than that of $\rho_{\Uparrow\Downarrow}$.
	
	$\rho_{\uparrow\Uparrow}$ and $\rho_{\uparrow\Downarrow}$ are defined as
	\begin{align}\label{sd3}
		\rho_{\uparrow\Uparrow}(k_x,k_y)&=f+ \frac{k_x}{m_{\rho}} g_T\,,\\
		\rho_{\uparrow\Downarrow}(k_x,k_y)&=f- \frac{k_x}{m_{\rho}}g_T\,,
	\end{align}
	$\rho_{\downarrow\Downarrow}=\rho_{\uparrow\Uparrow}$ and $\rho_{\downarrow\Uparrow}=\rho_{\uparrow\Downarrow}$. $\rho_{\uparrow\Uparrow}$ and $\rho_{\uparrow\Downarrow}$ represent the spin orientations of the quark and the $\rho$ meson, with longitudinal and transverse components, respectively. These are depicted in the third row of Fig. \ref{kaondp}. In the momentum space, $\rho_{\uparrow\Uparrow}$ and $\rho_{\uparrow\Downarrow}$ exhibit distortion effects due to the term $\frac{k_x}{m_{\rho}} g_T$. Specifically, for $\rho_{\uparrow\Uparrow}$, the axis of symmetry shifts to $(k_x\simeq0.09,k_y=0)$, while for $\rho_{\uparrow \Downarrow}$, it shifts to $(k_x\simeq-0.09,k_y=0)$.
	
	$\rho_{\Uparrow\uparrow}$ and $\rho_{\Downarrow\downarrow}$ are defined as
	\begin{align}\label{sd4}
		\rho_{\Uparrow\uparrow}(k_x,k_y)&=f+ \frac{k_x}{m_{\rho}} h_L^{\perp}\,,\\
		\rho_{\Uparrow\downarrow}(k_x,k_y)&=f- \frac{k_x}{m_{\rho}} h_L^{\perp}\,,
	\end{align}
	$\rho_{\Downarrow\downarrow}=\rho_{\Uparrow\uparrow}$ and $\rho_{\Downarrow\uparrow}=\rho_{\Uparrow\downarrow}$. $\rho_{\Downarrow\downarrow}=\rho_{\Uparrow\uparrow}$ and $\rho_{\Downarrow\uparrow}=\rho_{\Uparrow\downarrow}$. The quantities $\rho_{\Uparrow \uparrow}$ and $\rho_{\Downarrow \downarrow}$ are depicted in the final row of Fig. \ref{kaondp}, where they exhibit distortion effects due to the term $\frac{k_x}{m_\rho} h_L^{\perp}$. Specifically, for $\rho_{\Uparrow \uparrow}$, the axis of symmetry shifts to $(k_x\simeq-0.07,k_y=0)$, while for $\rho_{\Downarrow \downarrow}$, it moves to $(k_x\simeq0.07,k_y=0)$. A comparison between these two rows reveals that the distortion of a longitudinally-polarized quark in a transversely-polarized target is opposite to that of a transversely-polarized quark in a longitudinally polarized target. This phenomenon arises from the fact that $g_T$ is positive and $h_L^{\perp}$ is negative.
	
	

	\section{Summary and conclusion}\label{excellent}
	In this study, we assess the leading-twist transverse momentum dependent parton distributions (TMDs) of the $\rho$ meson using light-front wave functions (LFWFs) within the framework of the Nambu--Jona-Lasinio (NJL) model with proper time regularization.
	
	We have observed that the TMDs $f(x,\bm{k}_{\perp}^2)$, $g_L(x,\bm{k}_{\perp}^2)$, $g_T(x,\bm{k}_{\perp}^2)$, $h(x,\bm{k}_{\perp}^2)$ and $f_{TT}(x,\bm{k}_{\perp}^2)$ exhibit similar behavior, showing symmetry around $x=1/2$. As the transverse momentum $\bm{k}_{\perp}^2$ increases, these TMDs display a much weaker dependence on the light-cone momentum fraction $x$. This implies that a struck quark with large $\bm{k}_{\perp}^2$ has a much weaker preference for any particular value of $x$. For the longitudinal TMD $g_L(x,\bm{k}_{\perp}^2)$, the maximum value shifts to occur at  $x>1/2$. Conversely, for the transversely polarized TMD $h_L^{\perp}(x,\bm{k}_{\perp}^2)$, the minimum value corresponds to a momentum fraction $x<1/2$. In terms of tensor polarized TMDs, $f_{LL}(x,\bm{k}_{\perp}^2)$ and $f_{LT}(x,\bm{k}_{\perp}^2)$, we observe that $f_{LL}(x,\bm{k}_{\perp}^2)$ exhibits two peaks and is symmetric about $x=1/2$, while $f_{LT}(x,\bm{k}_{\perp}^2)$ is antisymmetric and equals zero at $x=1/2$ for all $\bm{k}_{\perp}^2$.
	
	We study the transverse momentum dependence of the $\rho$ meson TMDs, with a focus on their $\bm{k}_{\perp}$-weighted moments. In Ref.~\cite{Ninomiya:2017ggn}, the authors also calculate the $\rho$ meson TMDs in the NJL model using the covariant method. A comparison of $k_{\perp}$-weighted moments with their results shows that the numerical values are almost identical, except for $\langle k_{\perp}^n\rangle_{g_L}$, where our values are larger than theirs. Notably, our numerical values of $\langle k_{\perp}^n\rangle_{g_L}$ align with those of $\langle k_{\perp}^n\rangle_{h_L^{\perp}}$, consistent with findings in Refs.~\cite{Kaur:2020emh,Shi:2022erw}. Furthermore, we investigate the $x$-dependent average transverse momentum $\langle k_{\perp}^n(x)\rangle_{\alpha}$ of TMDs to determine typical transverse momenta of quark TMDs. Our results indicate that the typical transverse momenta of $\langle k_{\perp}(x)\rangle_{\alpha}$ fall within the range $[0.3, 0.45]$ GeV and those of $\langle k_{\perp}^2(x)\rangle_\alpha$ lie within $[0.15, 0.30]$ GeV$^2$ region.
	
	The four $\rho$ meson parton distribution function (PDFs) can be obtained by integrating the TMDs over $\bm{k}_{\perp}$. A comparison with Ref.~\cite{Ninomiya:2017ggn} reveals that $f(x)$, $f_{LL}(x)$, $h(x)$ and $h(x)$ are similar, but there is a difference in $g(x)$. In Ref.~\cite{Ninomiya:2017ggn}, $g(x)$ is negative when $x$ is small, whereas our results show that it is positive in the entire region of $x \in [0,1]$. We also conducted an investigation into the $x$-moments of various PDFs. Our spin sum rule indicates that $\Delta q= \langle x\rangle_{g} =0.505$, suggesting that the total valence quark and antiquark contribution to the spin of the $\rho$ meson is $50.5\%$, implying that about $49.5\%$ contribution comes from quark orbital angular momentum. The $x$-moments of $h(x)$ yield $\langle x\rangle_{h} =0.803$, indicating that the tensor charge at the model scale is 0.803.
	
	The positivity constraints of $\rho$ meson TMDs and PDFs are investigated in this study. The relationships $f_{LL}=-\frac{3}{2}f$, $|g|=\frac{3}{2}f$, and $|h|=\frac{M}{m_{\rho}}\frac{3}{2}f$ at the endpoints $x=0$ and $x=1$, as reported in Ref.~\cite{Ninomiya:2017ggn}, do not hold in our results, which is evident from Fig. \ref{pc}. The diagrams presented in Figs. \ref{pc} and \ref{tmdpc} demonstrate that our findings align well with the positivity constraints.
	
	Finally, we conducted a study on the spin densities within the $\rho$ meson in transverse momentum space. The spin densities are depicted in Fig. \ref{kaondp}. From the diagram, it is evident that both the quark and the target exhibit axial symmetry when polarized in either longitudinal or transverse directions. However, when the quark is longitudinally polarized and the target is transversely polarized (or vice versa), a dipolar distortion is observed in the symmetric distribution. Furthermore, opposite distortions are observed in configurations where the quark and $\rho$ meson are longitudinally-transversely or transversely-longitudinally polarized. Another discovery pertains to the quark spin densities of the $\rho$ meson, specifically the $\rho_{\uparrow \downarrow}\left(k_x,k_y\right)=f-\frac{1}{3}f_{LL}-g_L$. In the covariant approach, it is negative; however, $\rho_{\uparrow \downarrow}\left(k_x,k_y\right)$ from LFWFs is positive.
	
	For further investigation, our ongoing efforts to incorporate the effects of additional partons, such as sea quark contributions, may offer a more comprehensive understanding of the $\rho$ meson TMDs. Within the framework of the NJL model, these effects could be naturally represented by considering the influence of the virtual pion cloud surrounding the dressed quarks. Furthermore, we intend to account for higher Fock-state effects in future studies.
	
	\acknowledgments
	Work supported by: the Scientific Research Foundation of Nanjing Institute of Technology (Grant No. YKJ202352), the Natural Science Foundation of Jiangsu Province (Grant No. BK20191472), and the China Postdoctoral Science Foundation (Grant No. 2022M721564).

	\bibliographystyle{apsrev4-1}
	\bibliography{zhangrho}
	
	
\end{document}